\documentclass[sigconf, language=french,
language=german, language=spanish, language=english]{acmart}

\usepackage{xspace}
\usepackage{diagbox}
\usepackage{multirow}
\usepackage{lscape}

\newcommand{\Sim}{\textsf{Sim}\xspace}
\newcommand{\Util}{\textsf{Util}\xspace}

\newcommand{\zhe}[1]{{\color{red}Z:~#1}}
\newcommand{\ju}[1]{{\color{cyan}J:~#1}}

\newcommand{\tocheck}[1]{{\color{purple}#1}}

\copyrightyear{2023}
\acmYear{2023}
\setcopyright{rightsretained}
\acmConference[ICTIR '23]{Proceedings of the 2023 ACM SIGIR International Conference on the Theory of Information Retrieval}{July 23, 2023}{Taipei, Taiwan} \acmBooktitle{Proceedings of the 2023 ACM SIGIR International Conference on the Theory of Information Retrieval (ICTIR '23), July 23, 2023, Taipei, Taiwan}\acmDOI{10.1145/3578337.3605134}
\acmISBN{979-8-4007-0073-6/23/07}

\begin{document}

\title{Mitigating Mainstream Bias in Recommendation\\via Cost-sensitive Learning}

\author{Roger Zhe Li}
\email{Z.Li-9@tudelft.nl}
\affiliation{%
  \institution{Delft University of Technology}
  \city{Delft}
  \country{The Netherlands}
}

\author{Julián Urbano}
\email{J.Urbano@tudelft.nl}
\affiliation{%
  \institution{Delft University of Technology}
  \city{Delft}
  \country{The Netherlands}
}

\author{Alan Hanjalic}
\email{A.Hanjalic@tudelft.nl}
\affiliation{%
  \institution{Delft University of Technology}
  \city{Delft}
  \country{The Netherlands}
}

\renewcommand{\shorttitle}{Mitigating Mainstream Bias in Recommendation via Cost-sensitive Learning}

\begin{abstract}
Mainstream bias, where some users receive poor recommendations because their preferences are uncommon or simply because they are less active, is an important aspect to consider regarding fairness in recommender systems. Existing methods to mitigate mainstream bias do not explicitly model the importance of these non-mainstream users or, when they do, it is in a way that is not necessarily compatible with the data and recommendation model at hand. In contrast, we use the recommendation utility as a more generic and implicit proxy to quantify mainstreamness, and propose a simple user-weighting approach to incorporate it into the training process while taking the cost of potential recommendation errors into account. 
We provide extensive experimental results showing that quantifying mainstreamness via utility is better able at identifying non-mainstream users, and that they are indeed better served when training the model in a cost-sensitive way. This is achieved with negligible or no loss in overall recommendation accuracy, meaning that the models learn a better balance across users.
In addition, we show that research of this kind, which evaluates recommendation quality at the individual user level, may not be reliable if not using enough interactions when assessing model performance.
\end{abstract}
\sloppy

\ccsdesc[500]{Information systems~Recommender systems}

\keywords{Recommender Systems, Mainstream Bias, Bias Mitigation}



\maketitle

\section{Introduction} \label{sec:intro}
One of the critical limitations of recommender systems based on collaborative filtering (CF) models~\citep{DBLP:journals/cacm/GoldbergNOT92} is that they are \emph{not fair} in how they serve different groups of users~\cite{DBLP:conf/www/LeonhardtAK18, DBLP:conf/www/LiCFGZ21}. 
This fairness issue is a result of the varying quality of users' neighborhoods (groups of users with similar preferences) from which information is taken to train a CF model~\cite{DBLP:conf/wsdm/ZhuC22, DBLP:conf/wsdm/LiUH21}. The information collected from large, coherent, and information-rich neighborhoods will be the dominant one in steering the process of learning to recommend for all users. We refer to such dominant neighborhoods as \emph{mainstream}. Because the users belonging to such neighborhoods ---the \emph{mainstream users}--- are compatible with the learned model, they are optimally served.
For the \emph{non-mainstream} users, e.g. \emph{niche} groups who deviate from the mainstream and whose interaction information is therefore less rich~\citep{DBLP:conf/www/LiCFGZ21}, who are less active compared to the mainstream users~\cite{DBLP:conf/sigir/NaghiaeiRD22}, or where the preferences are not well pronounced, the neighborhoods cannot fully reflect their genuine preferences.
All this will make the non-mainstream users receive recommendations of a lower quality than the mainstream users. The difference in the quality of the CF model for these two user groups, further referred to as the \emph{mainstream bias}, will result in the continuous improvement of the performance for the mainstream group, and continuous decrease of the performance for the rest~\citep{DBLP:conf/kdd/LiuGZL19}.

While the issue of treating users differently by a recommender system in general has been addressed by a number of approaches, making for example assumptions about the relation between users' gender~\cite{DBLP:journals/ipm/MelchiorreRPBLS21} or demographics~\cite{DBLP:conf/fat/EkstrandTAEAMP18} and the quality of recommendation, not many approaches have focused specifically on addressing the mainstream bias. \citet{DBLP:conf/wsdm/LiUH21} deployed an autoencoder~\citep{rumelhart1985learning} for feature reconstruction as an adversary to a traditional CF model, forcing it to deviate from the pure similarity-based learning and make the learned model more compatible with the non-mainstream users. More specifically, the autoencoder was deployed to steer the process of learning the user/item representation space for rating prediction via optimal reconstruction of the properties of all users, mainstream and otherwise, assuming this would lead to equal treatment of users during recommendation. Still, a more explicit focus on the mainstreamness of users is needed to ensure that the bias is effectively addressed.

Inspired by outlier detection techniques, \citet{DBLP:conf/wsdm/ZhuC22} did focus on explicitly quantifying mainstreamness via similarities of user-preference profiles, and incorporated them to fine-tune the recommendation process for different user groups.
However, in the absence of ground truth data about mainstreamness, it is difficult to assess how well these approaches identify non-mainstream users. In addition, these mainstreamness statistics are model-agnostic in the sense that they are independent of the recommendation strategy, effectively ignoring the model's own capability to reduce the mainstream bias or even amplify it. As a result, the learning process could be tailored to the wrong users.

In this paper, we choose to focus there where the effect of mainstreamness is \emph{directly} observed, that is, the recommendation utility provided by the data and recommendation model at hand. 
If a user receives poor recommendations it could be because their preferences deviate from the rest, or because there is not enough data to properly quantify their similarity to other users or to fully exploit it. Therefore, we choose utility as an implicit proxy for mainstreamness.
Through this quantification of user mainstreamness, we make the training process focus on the non-mainstream ones by assigning them higher weights. 
We do so, however, 
in a \emph{cost-sensitive} way~\cite{DBLP:conf/ijcnn/Thai-NgheGS10}, taking the cost of recommendation errors into account while training the CF model.
Our results show that our implicit measurement of mainstreamness via utility is better able to differentiate niche users than an explicit approach, and that the cost-sensitive learning strategy does mitigate the bias by balancing the recommendation quality across users.
Finally, we investigate data requirements for conducting research on mainstream bias at the individual user level, and provide suggestions for reliable experimentation in this area.
\section{Proposed Approach}\label{sec:cost}

\begin{figure}
    \centering 
    \includegraphics[width=.23\textwidth]{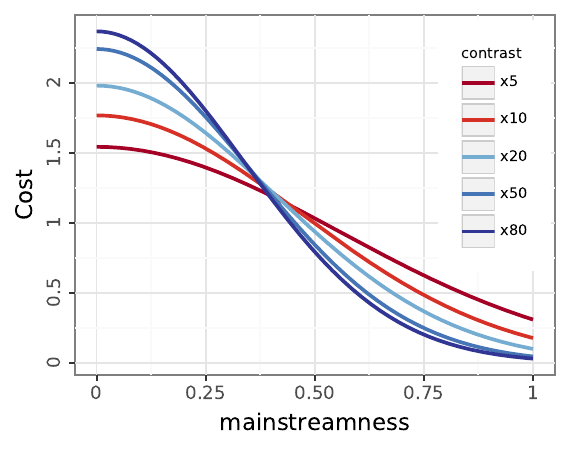}
    \caption{Cost functions used in the paper. The contrast denotes the relative cost between users with mainstreamness 0 and users with 1 (i.e. x10 means 10 times as much).}
    \label{fig:weight}
\end{figure}

The basis of our approach is a weighted loss function where every user $u\in\mathcal{U}$ is assigned a weight $\omega(u)$ that informs the learning process about the importance of every user's individual recommendation loss. The global loss is thus simply
\begin{equation}
\label{eq:overall}
        \mathcal{L} = \sum_{u\in\mathcal{U}}\omega(u) \mathcal{L}_R(u)~,
\end{equation}
where the recommendation loss $\mathcal{L}_R$ is specific of the model and learning paradigm. This way, we explicitly tell the learning process what users to optimize for by means of $\omega$, which, in our case, should be high for non-mainstream users and low for mainstream users.

\subsection{Definition of Weights} \label{subsec:def}

As explained in the previous section, we define $\omega$ as a function of the user mainstreamness $m_u$. 
However, rather than simply using a naïve transformation of $m_u$, we introduce flexibility through a cost function that maps user mainstreamness onto a cost value. In particular, and assuming $m_u$ ranges between 0 and 1, we use the density function of a Normal distribution truncated between 0 and 1, with zero mean and variance adjusted to achieve a contrast ranging between 5 (i.e. users with mainstreamness $m_u=0$ have a cost 5 times as large as users with $m_u=1$) and 80 (ie. 80 times as much). This is a simple choice to make $\omega$ smooth and monotonically decreasing, but other cost functions that emphasize different levels of mainstreamness are of course possible; we leave this discussion for further work. Fig.~\ref{fig:weight} shows some examples. 
Nonetheless, the formulation of the cost function may consider various aspects tailored to the business case, as well as different magnitudes for the contrast between users with low and high mainstreamness. For example, it would be reasonable to assign very high weights to non-mainstream users with high activity, or to users with very low activity as an attempt to reduce the churn rate.

An important point to consider when defining $\omega$ is the distribution of mainstreamness across users. It could be the case that, given the current data and model, the least mainstream users are actually fairly mainstream already, so their weight relative to the most mainstream users should be adjusted via a smaller contrast. 
It could also be the case that the dataset is very sparse and there are simply not enough neighbors around users for the model to learn a good representation. That is, the majority of users could be considered non-mainstream, and as a result the cost function would hardly differentiate among them. Lastly, one could decide to compute $m_u$ in several different ways (see next Section), which could potentially lead to quite different mainstreamness score distributions altogether, ultimately leading to a different set of weight values even for the same users.

In order to minimize this dependence on the dataset and mainstreamness definition, and ensure that the full co-domain of the cost function is used, we first normalize the raw mainstreamness scores.
Simply re-scaling between the minimum and maximum could still lead to a disproportionate use of small parts of the co-domain, and would also be very sensitive to outlier users. Instead, we use the rank statistic of $m_u$ normalized in $[0,1]$. We achieve this by using the empirical cumulative distribution function (ecdf)
\begin{equation}
    \omega(u)=\mathrm{cost}(\mathrm{ecdf}_{\mathcal{U}}(m_u))~,\label{eq:weight}
\end{equation}
where, as mentioned, cost is defined in terms of a truncated Normal density function.

\subsection{Measurement of Mainstreamness}

\begin{figure}[!t]
    \centering\includegraphics[scale=.5]{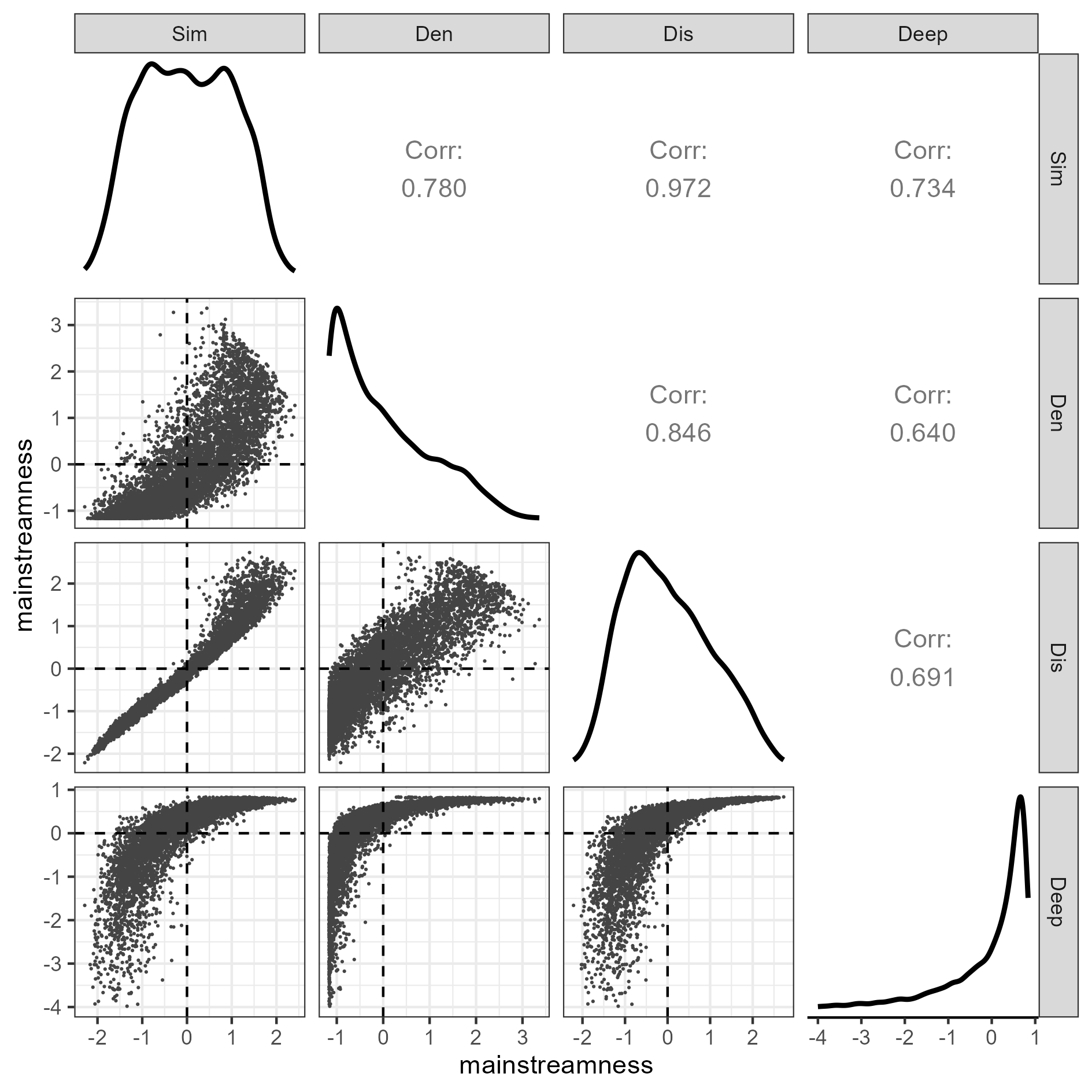}
    \caption{Comparison of the four mainstreamness definitions proposed by \citet{DBLP:conf/wsdm/ZhuC22}, as applied to the MovieLens 1M dataset.\protect\footnotemark Density plots illustrate the distributions of mainstreamness. Scatter plots show the relationship between pairs of definitions, quantified in the upper-right half via Pearson correlation scores. Scores are standardized to zero mean and unit variance for better comparison.}
    \label{fig:mvsm}
\end{figure}

\footnotetext{Data available from the authors' public repository at\\ \url{https://github.com/Zziwei/Measuring-Mitigating-Mainstream-Bias}.}

An \textbf{explicit} approach to compute $m_u$ would ideally follow some notion of mainstreamness, but mainstreamness is itself a complex construct very hard to define formally~\citep{DBLP:conf/ismir/0001S18, DBLP:conf/wsdm/ZhuC22, DBLP:conf/wsdm/LiUH21}. Recently, \citet{DBLP:conf/wsdm/ZhuC22} took inspiration from outlier detection techniques to propose four different definitions:
\begin{itemize}
    \item \textsf{Sim}: users are mainstream to the extent that their interactions are similar to that of the other users. The Jaccard coefficient is used to measure the average similarity between a user and all the others. 
    \item \textsf{Den}: users are mainstream to the extent that there are enough close neighbors to calculate similarity with. The local outlier factor algorithm (LOF)~\cite{DBLP:conf/sigmod/BreunigKNS00} is used to identify niche users.
    \item \textsf{Dis}: users are mainstream to the extent that their interactions are common in the dataset, that is, they interact with popular items. The cosine similarity is used to measure the similarity between a user and the average user interactions.
    \item \textsf{Deep}: similar to \textsf{Den}, niche users are identified by an outlier detection algorithm. In particular, the deep support vector data description algorithm (DeepSVDD)~\cite{DBLP:conf/icml/RuffGDSVBMK18} is used.
\end{itemize}

However, it is difficult to assess how well these, or any other definitions for that matter, correlate with the concept of mainstreamness. To illustrate, Fig.~\ref{fig:mvsm} compares these four definitions as applied to the MovieLens 1M dataset. Although they are somewhat correlated to one another, it is evident that they produce very different scores. For instance, \textsf{Sim} and \textsf{Dis} lead to nicely shaped distributions, suggesting few users with extreme (non-)mainstreamness. However, \textsf{Den} and \textsf{Deep} lead to very skewed distributions, even in the opposite direction, pointing to many users with extreme scores. This shows that the same user could be considered both mainstream or non-mainstream, depending on how we choose to define mainstreamness. 

Furthermore, it should be noted that these four definitions of mainstreamness are agnostic to the recommendation model. However, the effect of mainstreamness, ultimately, depends on the model and how it is able to exploit the specifics of the dataset it is trained on. It is not far-fetched to think of a user, assessed as non-mainstream, who receives bad recommendations under one model but good recommendations under a more capable one.

This leads us to consider an alternative, \textbf{implicit} way to quantify mainstreamness that is \textsl{not} model agnostic. In particular, we decide to focus there where the effect of mainstreamness is to be observed, that is, the recommendation \emph{utility} provided by the recommendation model at hand. This is where mainstreamness will ultimately have an impact on. The very nature of collaborative filtering tells us that if a user receives poor recommendations it is because they are non-mainstream under the current model: they cannot be properly represented, either because their preferences are somehow different from their closest neighbors, or because there are not enough data to properly quantify their similarity. Therefore, we use utility as a proxy for mainstreamness.
Since utility, just like mainstreamness, is a complex concept difficult to measure, we decide to simply use the accuracy of the recommendation model for that user, measured through a metric like $nDCG$ or $AP$. 

But there is the question of what accuracy scores we actually use. In principle, these scores should reflect user mainstreamness when there is no mechanism to minimize its effect, and they should be achieved by the recommendation model in the dataset at hand. Therefore, we decide to use the accuracy achieved, \textsl{on a validation set}, by the vanilla model whose loss function is as in Eq.~\eqref{eq:overall} but using no weights. As intended, we thus first see how the model reacts to mainstreamness as reflected in the observed utility for users, and then act upon it in a cost-sensitive way.

\section{Experimental Design} \label{sec:exp}

\begin{table}[t]
\caption{Dataset statistics after pre-filtering.}
    \centering{\small\setlength{\tabcolsep}{4pt}
    \begin{tabular}{@{}lrrrr@{}}
        \hline
        Dataset & \#users & \#items & \#ratings & Density \\
        \hline
        MovieLens 1M~\cite{DBLP:journals/tiis/HarperK16} & 6,040 & 3,609 & 562,957 & 2.583\% \\
        BeerAdvocate~\cite{DBLP:conf/icdm/McAuleyLJ12} & 8,821 & 43,663 & 780,752 & 0.203\% \\
        Amazon Digital Music~\cite{DBLP:conf/emnlp/NiLM19} & 14,057 & 379,171 & 619,673 & 0.011\% \\
        Amazon Musical Instruments ~\cite{DBLP:conf/emnlp/NiLM19} & 15,270 & 585,766 & 862,798 & 0.010\% \\
        \hline
      \end{tabular}}
    \label{tab:stats}
\end{table}

\begin{figure*}[!t]
    \centering\includegraphics[width=.5\columnwidth]{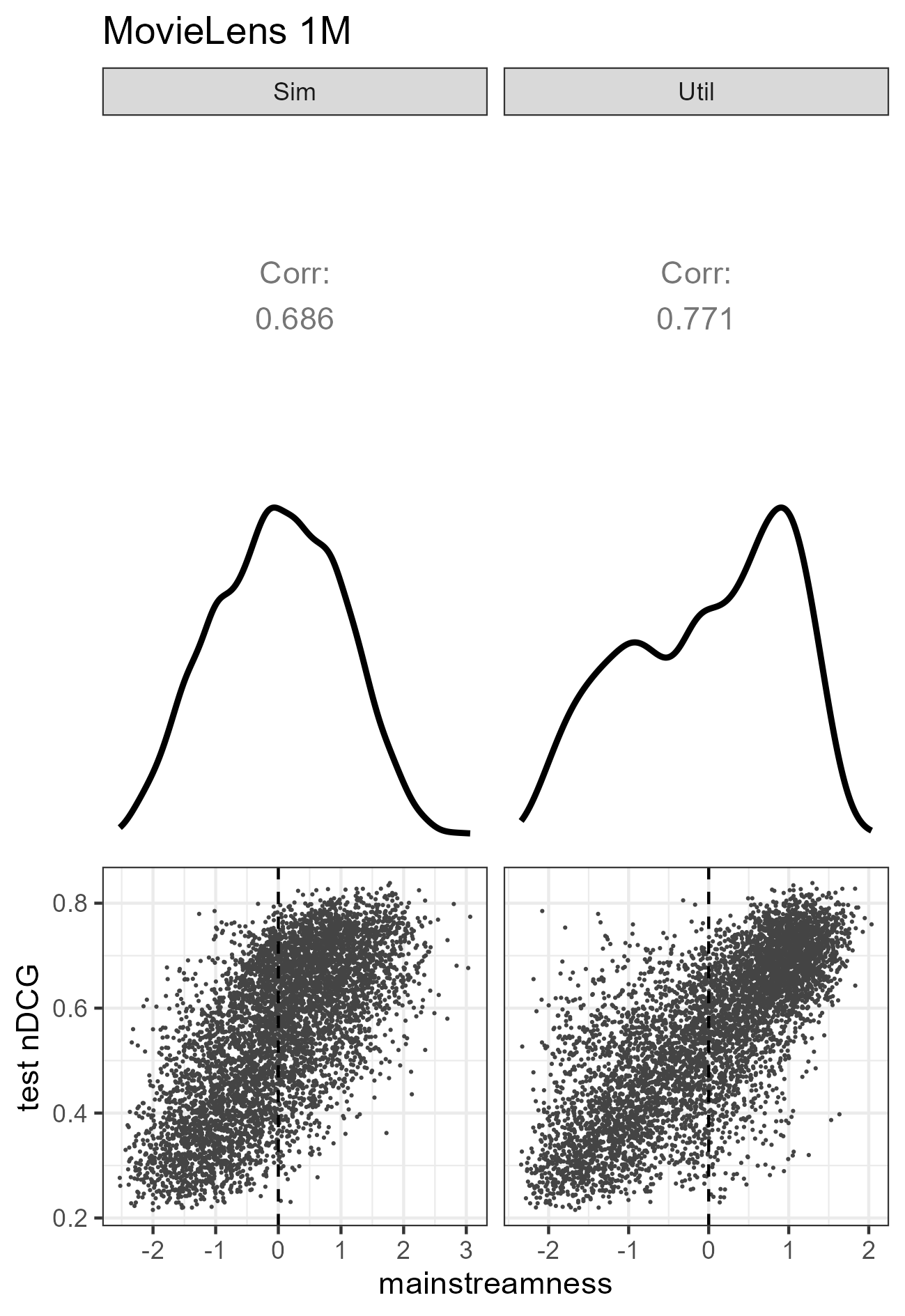}~\includegraphics[width=.5\columnwidth]{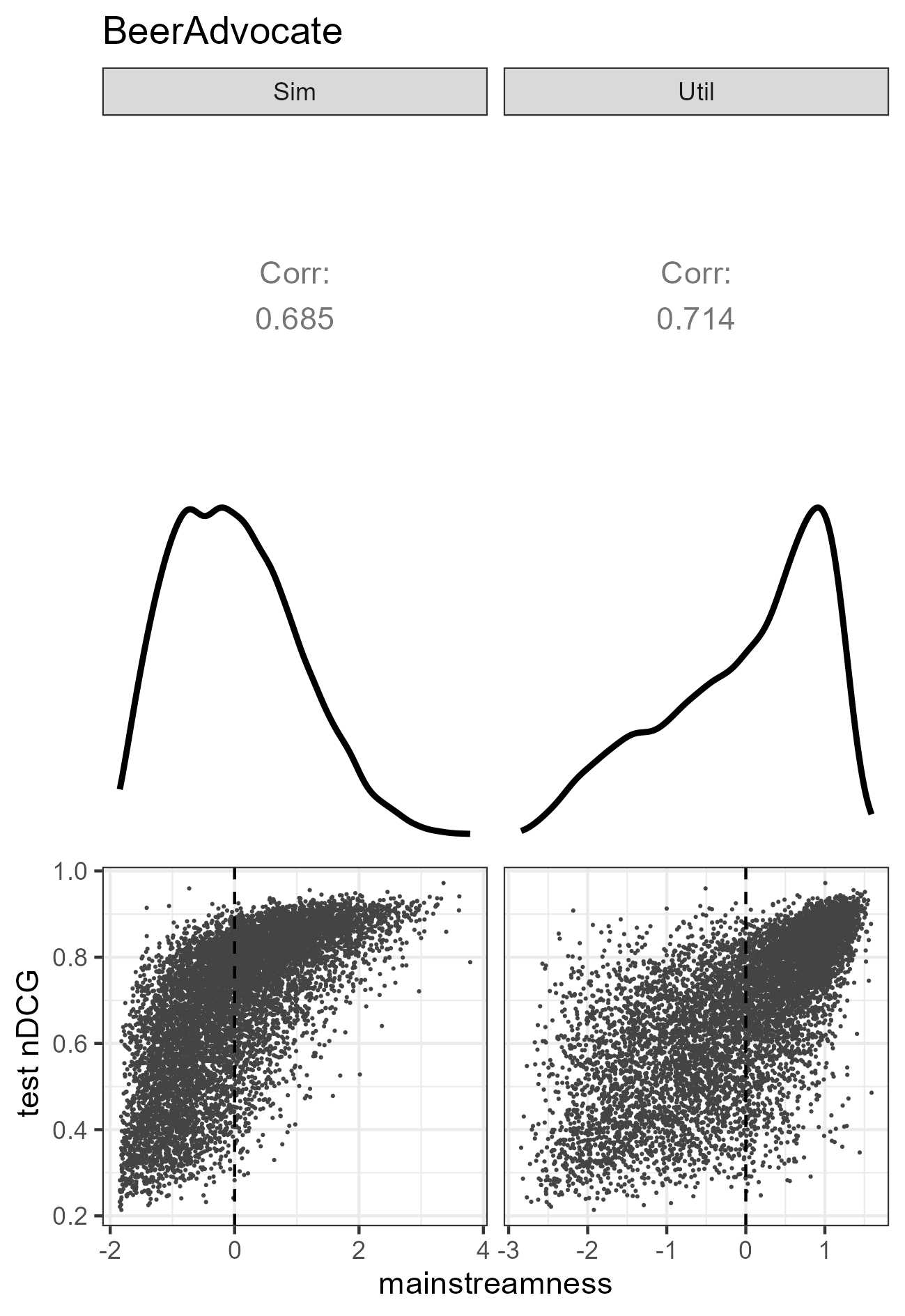}~\includegraphics[width=.5\columnwidth]{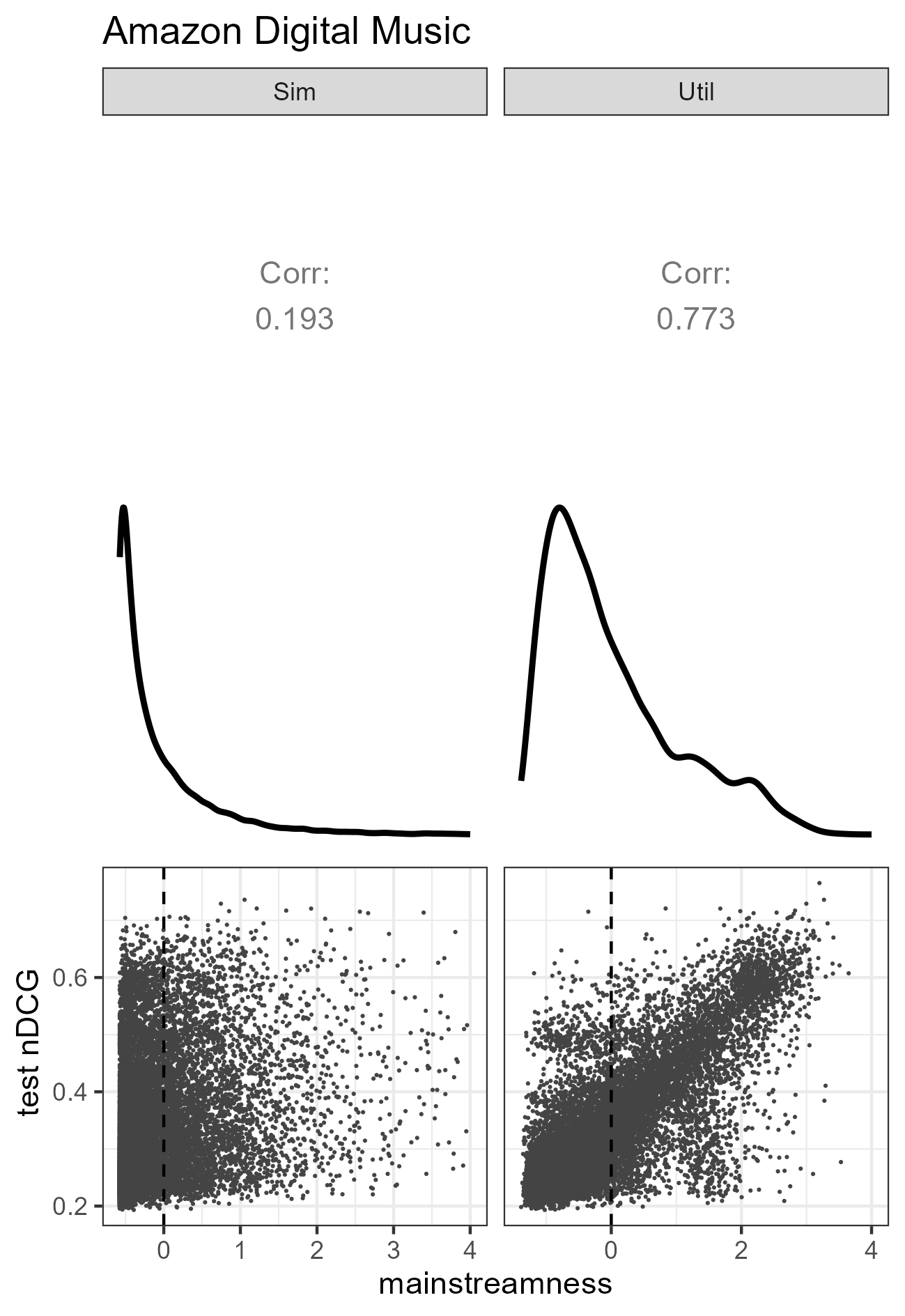}~\includegraphics[width=.5\columnwidth]{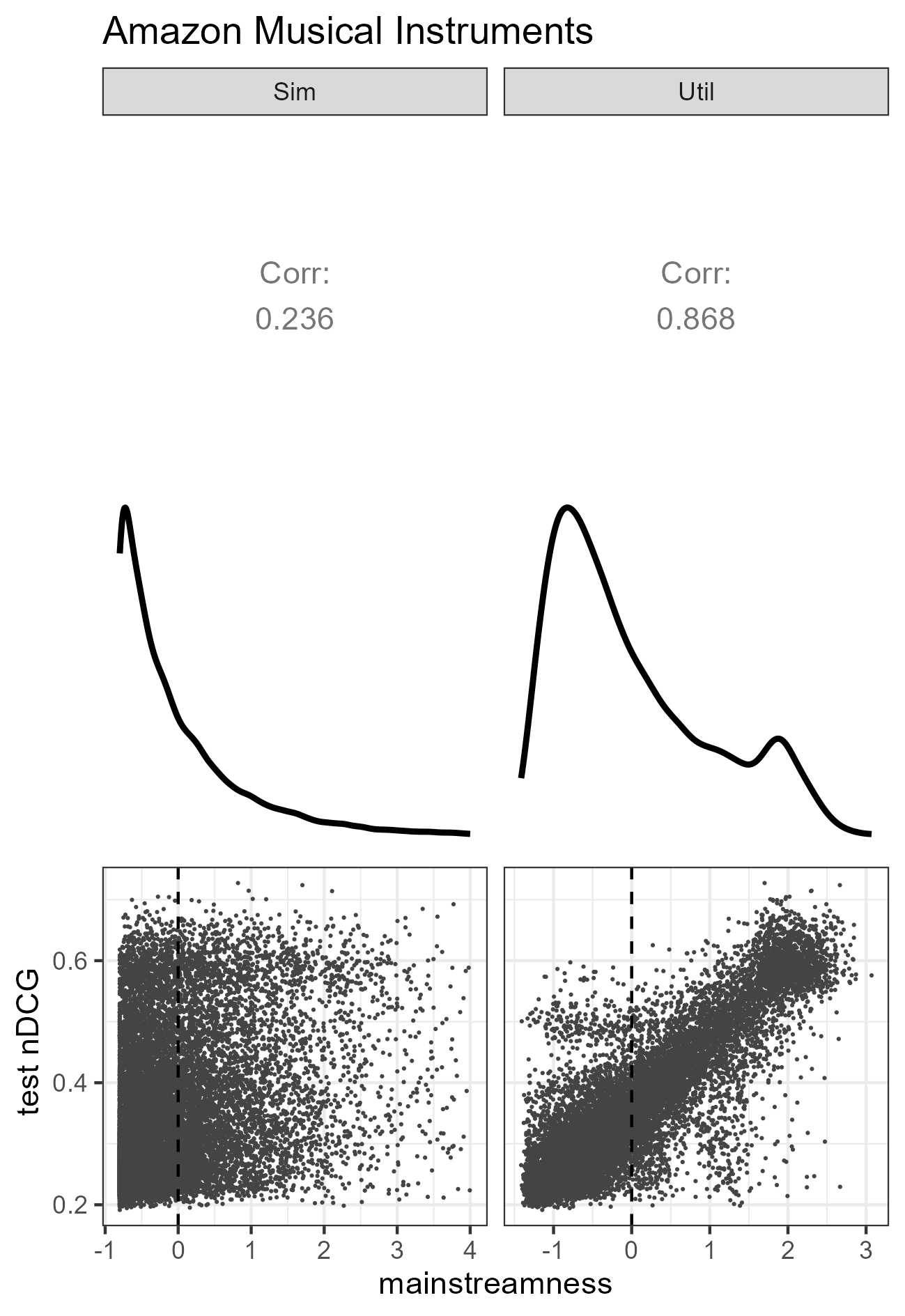}
    \caption{Correlation between mainstreamness and test nDCG in the baseline model (FM), for each  mainstreamness definition. Density plots illustrate the distribution of mainstreamness. Scatterplots show their relationship with nDCG, quantified via Pearson correlation scores. Mainstreamness scores are standardized to zero mean and unit variance for better comparison.}
    \label{fig:simvsutil}
\end{figure*}

\begin{figure*}[!t]
    \centering\includegraphics[scale=.51]{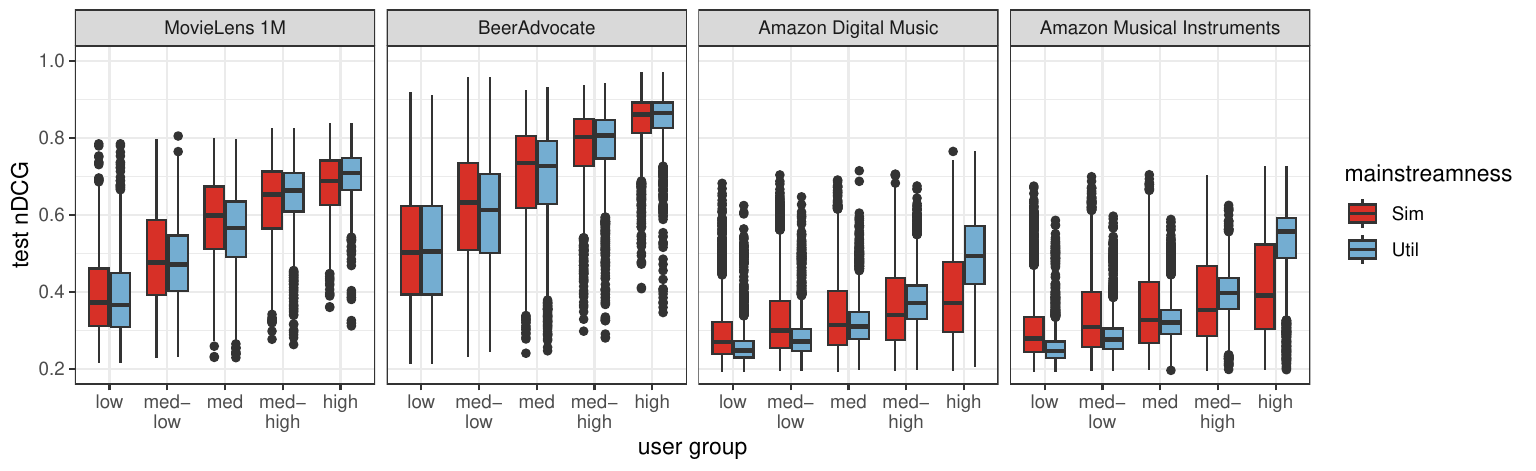}
    \caption{Correlation between user groups, split by mainstreamness, and test nDCG in the baseline model (FM).}
    \label{fig:boxes}
\end{figure*}

We carried out a number of experiments to investigate the effectiveness of the proposed approach in mitigating the mainstreamness bias, as well as the effect of the contrast applied by the cost function.
In particular, we study contrasts x5, x10, x20, x50 and x80, that is, the most non-mainstream user has a weight between 5 and 80 times larger than that of the most mainstream user. Fig.~\ref{fig:weight} details the cost functions. Regarding the measurement of mainstreamness, we consider both an explicit and an implicit quantification. For the former, we follow \citet{DBLP:conf/wsdm/ZhuC22} and compute \textsf{Sim} scores. This choice is motivated by the time complexity of their four approaches (the computation of mainstreamness may quickly become intractable as the numbers of users and items increase; while their datasets include a few thousand items, ours span from a few thousands to over half a million), and their correlation to one another (\textsf{Sim} is also the one most correlated with the others, in particular with \textsf{Deep}). For the implicit quantification we compute utility scores using the metric $nDCG$ as an exemplar of recommender systems research; hereafter, we will refer to this definition of mainstreamness as \textsf{Util}. 

We selected four real-world datasets containing user-item rating interactions from various domains and with different densities, especially including some highly sparse datasets (see Table~\ref{tab:stats}). In line with common practice in ranking-oriented recommender systems research, we see all existing interactions in the datasets as relevant, and all other unseen interactions as irrelevant. 
We use LensKit~\cite{DBLP:conf/cikm/Ekstrand20} to evenly split the relevant items for each user into training, validation and test sets. To make the modeling of utility ---and hence mainstreamness--- robust, each user has at least five relevant transactions in each of the three sets; we explain the rationale for this decision in Section~\ref{sec:discussion}.
For training the model, we follow \citet{DBLP:conf/www/HeLZNHC17, DBLP:conf/ijcai/WuWH22} and randomly sample four irrelevant items per relevant item in the training partition. 
For validation and test, we follow DaisyRec~\cite{DBLP:conf/recsys/SunY00Q0G20} and evaluate the model for each user by ranking a total of 500 items consisting of their relevant items in the validation/test partition and a set of randomly sampled irrelevant items.
Finally, to make sure relevant items are the minority, as happens in reality, we truncate the number of relevant interactions to 200.
The dataset statistics after processing are shown in Table~\ref{tab:stats}. 

Regarding the recommendation model, we deploy a simple but effective CF model that only utilizes user-item interactions.
Specifically, we choose Factorization Machines (FM) \cite{DBLP:conf/icdm/Rendle10}, which optimize the binary cross-entropy (BCE) loss via the Adaptive Moment Estimation (Adam)~\cite{DBLP:journals/corr/KingmaB14} learner, and leave the investigation on other training paradigms for future work. For each user, the BCE loss is normalized by dividing by the total number of relevant and irrelevant items used for training, so that all user losses are on the same scale in~\eqref{eq:overall}. 
After a fine-tuning process based on grid search, we fixed several key hyper-parameters including the dimension of vectors used for interaction (32), learning rate (0.0001), L2-regularization coefficient to avoid overfitting (0.001), and batch size (512). 

All models are trained for 300 epochs to ensure full convergence, and with 3 different random initializations to minimize random effects due to the sampling process. The whole pipeline is implemented in PyTorch \cite{DBLP:conf/nips/PaszkeGMLBCKLGA19}, and all experiments are run on one NVIDIA GeForce GTX 2080Ti GPU \footnote{All data, code and results are available at\\\url{https://github.com/roger-zhe-li/ictir23-cost-sensitive}.}. 

\newcommand{\sig}[1]{{#1}}
\begin{table*}[!th]
\caption{Mean nDCG of the baseline model (FM) per user group, and relative percentage improvement of each cost-sensitive model (e.g. users in group `low' of MovieLens 1M received a score of .3284 with the baseline, and an improvement of +3.89\% with the x80-contrast cost-sensitive model under the \textsf{Util} mainstreamness definition). Column `Overall' lists the mean across all users. Green/red for statistically significant gain/loss with respect to the baseline (hierarchical linear model with seed and user random effects, Bonferroni correction).}\label{tab:overall_performance}
\footnotesize\setlength{\tabcolsep}{1.9pt}
\begin{tabular}{|cr|r|lllll|r|lllll|r|lllll|r|lllll|}
\cline{3-26}
\multicolumn{2}{c|}{} & \multicolumn{6}{c|}{MovieLens 1M} & \multicolumn{6}{c|}{BeerAdvocate} & \multicolumn{6}{c|}{Amazon Digital Music} & \multicolumn{6}{c|}{Amazon Musical Instruments} \\
\cline{3-26}
\multicolumn{2}{c|}{} & & & med- & & med- & & & & med- & & med- & & & & med- & & med- & & & & med- & & med- & \\
\multicolumn{2}{c|}{} & Overall & low & low & med & high & high & Overall & low & low & med & high & high & Overall & low & low & med & high & high & Overall & low & low & med & high & high \\
\hline

\multicolumn{2}{|c|}{FM} & .5531 & .3284 & .4621 & .5753 & .6613 & .7388 & .6887 & .4144 & .6051 & .7301 & .8132 & .8809 & .3456 & .2324 & .2695 & .3145 & .3828 & .5289 & .3606 & .2348 & .2772 & .3276 & .4085 & .5552 \\ \hline

\multirow{5}{*}{\rotatebox[origin=c]{90}{Sim}} & x5  & {\color[HTML]{9A0000} \sig{.5465}} & {-0.36}       & {\color[HTML]{9A0000} \sig{-0.89}} & {\color[HTML]{9A0000} \sig{-1.62}} & {\color[HTML]{9A0000} \sig{-1.3}}  & {\color[HTML]{9A0000} \sig{-1.33}} & {\color[HTML]{9A0000} \sig{.6792}} & {-0.52}       & {\color[HTML]{9A0000} \sig{-1.72}} & {\color[HTML]{9A0000} \sig{-1.83}} & {\color[HTML]{9A0000} \sig{-1.44}} & {\color[HTML]{9A0000} \sig{-1.13}} & {\color[HTML]{9A0000} \sig{.3395}} & {\color[HTML]{009901} \sig{+0.65}} & {+0.05}       & {\color[HTML]{9A0000} \sig{-0.61}} & {\color[HTML]{9A0000} \sig{-1.8}}  & {\color[HTML]{9A0000} \sig{-4.45}}  & {\color[HTML]{9A0000} \sig{.3581}} & {\color[HTML]{009901} \sig{+1.61}} & {\color[HTML]{009901} \sig{+0.77}} & {-0.09}       & {\color[HTML]{9A0000} \sig{-1.06}} & {\color[HTML]{9A0000} \sig{-2.5}}  \\
 & x10 & {\color[HTML]{9A0000} \sig{.5437}} & {-0.47}       & {\color[HTML]{9A0000} \sig{-1.32}} & {\color[HTML]{9A0000} \sig{-2.23}} & {\color[HTML]{9A0000} \sig{-1.87}} & {\color[HTML]{9A0000} \sig{-1.94}} & {\color[HTML]{9A0000} \sig{.6734}} & {\color[HTML]{9A0000} \sig{-0.81}} & {\color[HTML]{9A0000} \sig{-2.87}} & {\color[HTML]{9A0000} \sig{-2.94}} & {\color[HTML]{9A0000} \sig{-2.27}} & {\color[HTML]{9A0000} \sig{-1.8}}  & {\color[HTML]{9A0000} \sig{.3368}} & {\color[HTML]{009901} \sig{+0.99}} & {+0.1}        & {\color[HTML]{9A0000} \sig{-1}}    & {\color[HTML]{9A0000} \sig{-2.67}} & {\color[HTML]{9A0000} \sig{-6.27}}  & {\color[HTML]{9A0000} \sig{.3577}} & {\color[HTML]{009901} \sig{+3.29}} & {\color[HTML]{009901} \sig{+1.44}} & {+0.08}       & {\color[HTML]{9A0000} \sig{-1.68}} & {\color[HTML]{9A0000} \sig{-3.54}} \\
 & x20 & {\color[HTML]{9A0000} \sig{.541}}  & {-0.53}       & {\color[HTML]{9A0000} \sig{-1.76}} & {\color[HTML]{9A0000} \sig{-2.85}} & {\color[HTML]{9A0000} \sig{-2.41}} & {\color[HTML]{9A0000} \sig{-2.51}} & {\color[HTML]{9A0000} \sig{.6666}} & {\color[HTML]{9A0000} \sig{-1.35}} & {\color[HTML]{9A0000} \sig{-4.23}} & {\color[HTML]{9A0000} \sig{-4.13}} & {\color[HTML]{9A0000} \sig{-3.22}} & {\color[HTML]{9A0000} \sig{-2.56}} & {\color[HTML]{9A0000} \sig{.3347}} & {\color[HTML]{009901} \sig{+1.31}} & {+0.09}       & {\color[HTML]{9A0000} \sig{-1.3}}  & {\color[HTML]{9A0000} \sig{-3.46}} & {\color[HTML]{9A0000} \sig{-7.69}}  & {\color[HTML]{9A0000} \sig{.3576}} & {\color[HTML]{009901} \sig{+3.33}} & {\color[HTML]{009901} \sig{+1.44}} & {+0.04}       & {\color[HTML]{9A0000} \sig{-1.72}} & {\color[HTML]{9A0000} \sig{-3.58}} \\
 & x50 & {\color[HTML]{9A0000} \sig{.5376}} & {-0.67}       & {\color[HTML]{9A0000} \sig{-2.28}} & {\color[HTML]{9A0000} \sig{-3.64}} & {\color[HTML]{9A0000} \sig{-3.06}} & {\color[HTML]{9A0000} \sig{-3.26}} & {\color[HTML]{9A0000} \sig{.6588}} & {\color[HTML]{9A0000} \sig{-2.01}} & {\color[HTML]{9A0000} \sig{-5.65}} & {\color[HTML]{9A0000} \sig{-5.56}} & {\color[HTML]{9A0000} \sig{-4.29}} & {\color[HTML]{9A0000} \sig{-3.54}} & {\color[HTML]{9A0000} \sig{.3328}} & {\color[HTML]{009901} \sig{+1.63}} & {+0.07}       & {\color[HTML]{9A0000} \sig{-1.6}}  & {\color[HTML]{9A0000} \sig{-4.13}} & {\color[HTML]{9A0000} \sig{-8.92}}  & {\color[HTML]{9A0000} \sig{.3576}} & {\color[HTML]{009901} \sig{+3.37}} & {\color[HTML]{009901} \sig{+1.45}} & {+0.03}       & {\color[HTML]{9A0000} \sig{-1.75}} & {\color[HTML]{9A0000} \sig{-3.6}}  \\
 & x80 & {\color[HTML]{9A0000} \sig{.5359}} & {-0.67}       & {\color[HTML]{9A0000} \sig{-2.55}} & {\color[HTML]{9A0000} \sig{-4.06}} & {\color[HTML]{9A0000} \sig{-3.39}} & {\color[HTML]{9A0000} \sig{-3.59}} & {\color[HTML]{9A0000} \sig{.6548}} & {\color[HTML]{9A0000} \sig{-2.55}} & {\color[HTML]{9A0000} \sig{-6.39}} & {\color[HTML]{9A0000} \sig{-6.17}} & {\color[HTML]{9A0000} \sig{-4.84}} & {\color[HTML]{9A0000} \sig{-4.08}} & {\color[HTML]{9A0000} \sig{.3316}} & {\color[HTML]{009901} \sig{+3.66}} & {\color[HTML]{009901} \sig{+0.9}}  & {\color[HTML]{9A0000} \sig{-1.78}} & {\color[HTML]{9A0000} \sig{-5.07}} & {\color[HTML]{9A0000} \sig{-10.62}} & {\color[HTML]{9A0000} \sig{.3576}} & {\color[HTML]{009901} \sig{+3.39}} & {\color[HTML]{009901} \sig{+1.45}} & {+0.02}       & {\color[HTML]{9A0000} \sig{-1.77}} & {\color[HTML]{9A0000} \sig{-3.61}} \\ \hline

\multirow{5}{*}{\rotatebox[origin=c]{90}{Util}} & x5  & {\color[HTML]{009901} \sig{.5567}} & {\color[HTML]{009901} \sig{+1.67}} & {\color[HTML]{009901} \sig{+1.81}} & {\color[HTML]{009901} \sig{+0.63}} & {+0.16}       & {-0.13}       & {\color[HTML]{9A0000} \sig{.6846}} & {+0.63}       & {\color[HTML]{9A0000} \sig{-0.41}} & {\color[HTML]{9A0000} \sig{-0.94}} & {\color[HTML]{9A0000} \sig{-0.83}} & {\color[HTML]{9A0000} \sig{-0.77}} & {.3454}       & {\color[HTML]{009901} \sig{+1.59}} & {\color[HTML]{009901} \sig{+1.43}} & {\color[HTML]{009901} \sig{+1.2}}  & {\color[HTML]{009901} \sig{+0.4}}  & {\color[HTML]{9A0000} \sig{-2.58}}  & {.3607}       & {\color[HTML]{009901} \sig{+1.1}}  & {\color[HTML]{009901} \sig{+0.96}} & {\color[HTML]{009901} \sig{+0.62}} & {-0.04}       & {\color[HTML]{9A0000} \sig{-1.19}} \\
 & x10 & {\color[HTML]{009901} \sig{.5574}} & {\color[HTML]{009901} \sig{+2.38}} & {\color[HTML]{009901} \sig{+2.34}} & {\color[HTML]{009901} \sig{+0.7}}  & {+0.11}       & {\color[HTML]{9A0000} \sig{-0.27}} & {\color[HTML]{9A0000} \sig{.6807}} & {+0.44}       & {\color[HTML]{9A0000} \sig{-1.19}} & {\color[HTML]{9A0000} \sig{-1.66}} & {\color[HTML]{9A0000} \sig{-1.39}} & {\color[HTML]{9A0000} \sig{-1.27}} & {.3453}       & {\color[HTML]{009901} \sig{+2.54}} & {\color[HTML]{009901} \sig{+2.09}} & {\color[HTML]{009901} \sig{+1.53}} & {+0.18}       & {\color[HTML]{9A0000} \sig{-3.5}}   & {.3607}       & {\color[HTML]{009901} \sig{+2}}    & {\color[HTML]{009901} \sig{+1.52}} & {\color[HTML]{009901} \sig{+0.78}} & {-0.32}       & {\color[HTML]{9A0000} \sig{-1.8}}  \\
 & x20 & {\color[HTML]{009901} \sig{.5579}} & {\color[HTML]{009901} \sig{+3.05}} & {\color[HTML]{009901} \sig{+2.87}} & {\color[HTML]{009901} \sig{+0.73}} & {0}           & {\color[HTML]{9A0000} \sig{-0.48}} & {\color[HTML]{9A0000} \sig{.6762}} & {+0.54}       & {\color[HTML]{9A0000} \sig{-2.11}} & {\color[HTML]{9A0000} \sig{-2.67}} & {\color[HTML]{9A0000} \sig{-2.09}} & {\color[HTML]{9A0000} \sig{-1.74}} & {.3454}       & {\color[HTML]{009901} \sig{+3.59}} & {\color[HTML]{009901} \sig{+2.78}} & {\color[HTML]{009901} \sig{+1.64}} & {+0.11}       & {\color[HTML]{9A0000} \sig{-4.25}}  & {.3607}       & {\color[HTML]{009901} \sig{+2.63}} & {\color[HTML]{009901} \sig{+1.93}} & {\color[HTML]{009901} \sig{+0.8}}  & {\color[HTML]{9A0000} \sig{-0.53}} & {\color[HTML]{9A0000} \sig{-2.08}} \\
 & x50 & {\color[HTML]{009901} \sig{.5579}} & {\color[HTML]{009901} \sig{+3.62}} & {\color[HTML]{009901} \sig{+3.31}} & {\color[HTML]{009901} \sig{+0.68}} & {-0.21}       & {\color[HTML]{9A0000} \sig{-0.84}} & {\color[HTML]{9A0000} \sig{.6722}} & {\color[HTML]{009901} \sig{+2}}    & {\color[HTML]{9A0000} \sig{-3.09}} & {\color[HTML]{9A0000} \sig{-3.86}} & {\color[HTML]{9A0000} \sig{-2.89}} & {\color[HTML]{9A0000} \sig{-2.32}} & {.3458}       & {\color[HTML]{009901} \sig{+4.84}} & {\color[HTML]{009901} \sig{+3.67}} & {\color[HTML]{009901} \sig{+1.92}} & {-0.11}       & {\color[HTML]{9A0000} \sig{-4.9}}   & {.3608}       & {\color[HTML]{009901} \sig{+3.94}} & {\color[HTML]{009901} \sig{+2.48}} & {\color[HTML]{009901} \sig{+0.85}} & {\color[HTML]{9A0000} \sig{-0.83}} & {\color[HTML]{9A0000} \sig{-2.66}} \\
 & x80 & {\color[HTML]{009901} \sig{.5577}} & {\color[HTML]{009901} \sig{+3.89}} & {\color[HTML]{009901} \sig{+3.47}} & {\color[HTML]{009901} \sig{+0.63}} & {\color[HTML]{9A0000} \sig{-0.32}} & {\color[HTML]{9A0000} \sig{-1.05}} & {\color[HTML]{9A0000} \sig{.6715}} & {\color[HTML]{009901} \sig{+2.98}} & {\color[HTML]{9A0000} \sig{-3.2}}  & {\color[HTML]{9A0000} \sig{-4.25}} & {\color[HTML]{9A0000} \sig{-3.14}} & {\color[HTML]{9A0000} \sig{-2.54}} & {.346}        & {\color[HTML]{009901} \sig{+5.45}} & {\color[HTML]{009901} \sig{+4}}    & {\color[HTML]{009901} \sig{+2.02}} & {-0.18}       & {\color[HTML]{9A0000} \sig{-5.15}}  & {.3608}       & {\color[HTML]{009901} \sig{+4.48}} & {\color[HTML]{009901} \sig{+2.64}} & {\color[HTML]{009901} \sig{+0.88}} & {\color[HTML]{9A0000} \sig{-0.97}} & {\color[HTML]{9A0000} \sig{-2.86}} \\ \hline
\end{tabular}
\end{table*}

\begin{figure}[t]
    \centering 
    \includegraphics[width=\columnwidth]{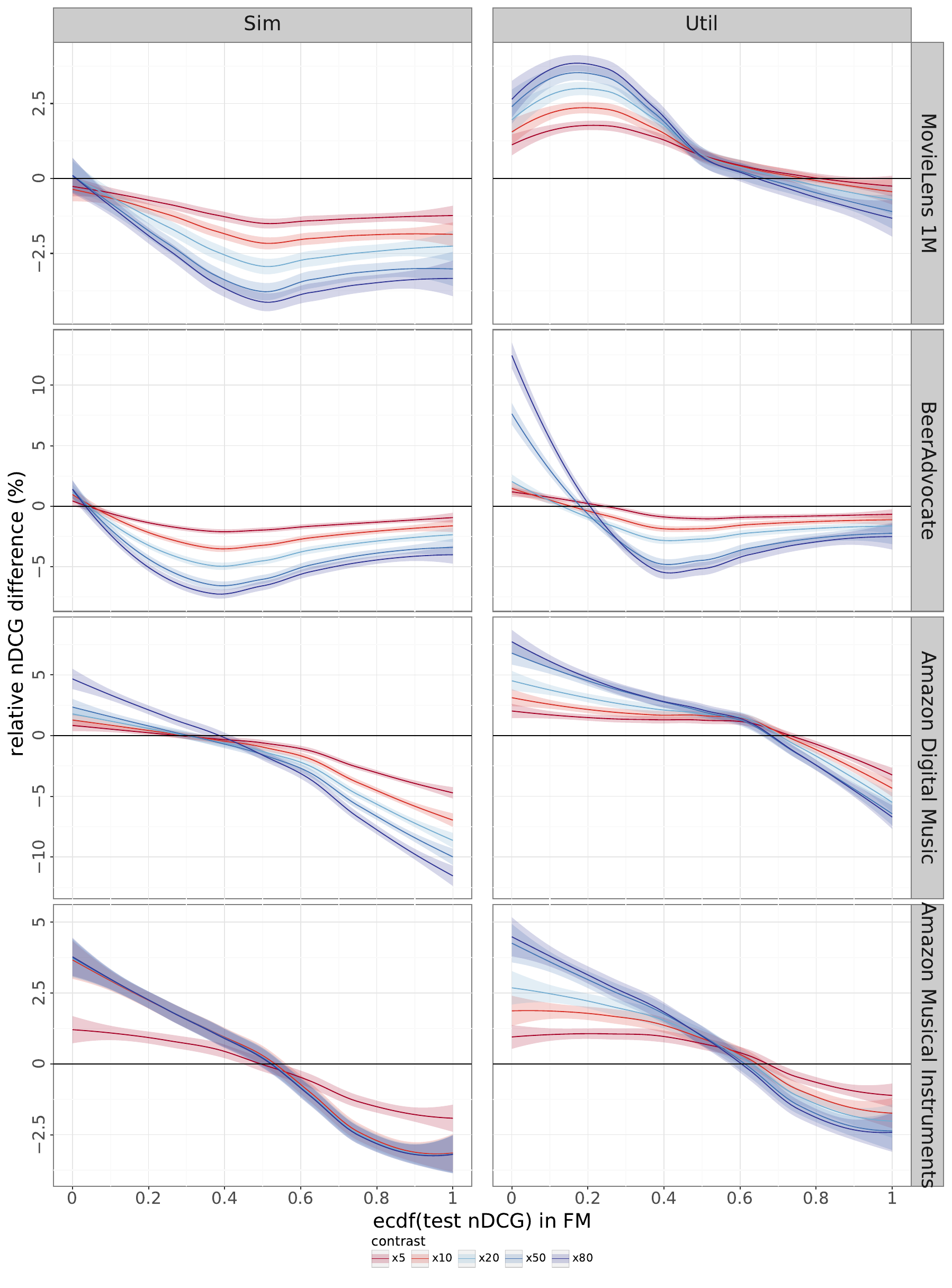}
    \caption{Mean nDCG relative percentage improvement between cost-sensitive models and baseline model, as a function ecdf(test nDCG) in the baseline FM model, for a sample data split. Curves fitted by a LOESS model. Ribbons indicate 95\% confidence intervals.}
    \label{fig:res_csl_by_ndcg}
\end{figure}

\section{Results} \label{sec:res}


\subsection{Mainstreamness and Utility}\label{subsec:m_and_u}

We first examine how \Sim and \Util differentiate between mainstream and non-mainstream users. In particular, we are interested in how well they correlate with the test $nDCG$ scores obtained by the baseline FM model: non-mainstream users should receive recommendations with low $nDCG$ scores, while mainstream users should receive higher scores.

For each of the four datasets, Fig.~\ref{fig:simvsutil} compares \Sim and \Util. We can first see that both approaches lead to similar distributions in the Amazon datasets, where there appear to be many non-mainstream users. However, they somewhat disagree in the BeerAdvocate dataset, where \Util does not identify many non-mainstream users to benefit from the cost-sensitive approach. In terms of correlation with the test $nDCG$ scores, we can see that \Util is much better correlated, specially in the Amazon datasets. This points to the possibility that \Sim identifies many non-mainstream users to which the model is still able to offer good recommendations. If the training process increases their importance by assigning them a high weight $\omega$, we may loose the opportunity to focus on those users that still receive poor recommendations.

In order to assess the effectiveness of the cost-sensitive approach for the mitigation of the mainstream bias, we will look in the next Section into different groups of users separated by their mainstreamness: group `low' contains the 20\% of users with lowest mainstreamness scores on the baseline model, group `med-low' contains the next 20\% or users, group `med' contains the middle 20\% of users, and so on with groups `med-high' and `high'. An effective mitigation of the mainstream bias would be reflected in increased performance for the lower groups, which ideally should be those with lowest test $nDCG$ scores in the baseline model.
Fig.~\ref{fig:boxes} shows how well \Sim and \Util separate users in these five groups. We can first see that the groups are indeed correlated with $nDCG$, but we can notice that this correlation is stronger with \Util, specially in the Amazon datasets (the low groups receive lower utility, and the higher groups receive higher utility). We can also see that groups tend to overlap substantially when separated by \Sim, potentially misplacing users. This overlap can be quantified by an ANOVA model of $nDCG$ modeled by two factors: dataset and user-group nested within dataset. Indeed, the user-group effect has a much larger sum of squares (SS) with \Util than with \Sim (SS=440 vs SS=218; SS of the dataset effect is 843).
Finally, Fig.~\ref{fig:boxes} also points that the BeerAdvocate dataset may be hard to further optimize for because the utility scores are already relatively high.

\subsection{Bias Mitigation}\label{ssec:mitigation}

\begin{table*}[!th]
\caption{Same as Table~\ref{tab:overall_performance}, but user groups defined by \Sim scores instead of test nDCG in the baseline model.}\label{tab:overall_performance_by_ms}
\footnotesize\setlength{\tabcolsep}{1.9pt}
\begin{tabular}{|cr|r|lllll|r|lllll|r|lllll|r|lllll|}
\cline{3-26}
\multicolumn{2}{c|}{} & \multicolumn{6}{c|}{MovieLens 1M} & \multicolumn{6}{c|}{BeerAdvocate} & \multicolumn{6}{c|}{Amazon Digital Music} & \multicolumn{6}{c|}{Amazon Musical Instruments} \\
\cline{3-26}
\multicolumn{2}{c|}{} & & & med- & & med- & & & & med- & & med- & & & & med- & & med- & & & & med- & & med- & \\
\multicolumn{2}{c|}{} & Overall & low & low & med & high & high & Overall & low & low & med & high & high & Overall & low & low & med & high & high & Overall & low & low & med & high & high \\
\hline

\multicolumn{2}{|c|}{FM} & .5531 & .3284 & .4621 & .5753 & .6613 & .7388 & .6887 & .4144 & .6051 & .7301 & .8132 & .8809 & .3456 & .2324 & .2695 & .3145 & .3828 & .5289 & .3606 & .2348 & .2772 & .3276 & .4085 & .5552 \\ \hline

\multirow{5}{*}{\rotatebox[origin=c]{90}{Sim}} & x5  & {\color[HTML]{9A0000} \sig{.5465}} & -0.62                              & -0.97                              & {\color[HTML]{9A0000} \sig{-1.16}} & {\color[HTML]{9A0000} \sig{-1.34}} & {\color[HTML]{9A0000} \sig{-1.58}} & {\color[HTML]{9A0000} \sig{.6792}} & {\color[HTML]{9A0000} \sig{-1.4}}  & {\color[HTML]{9A0000} \sig{-1.39}} & {\color[HTML]{9A0000} \sig{-1.43}} & {\color[HTML]{9A0000} \sig{-1.34}} & {\color[HTML]{9A0000} \sig{-1.35}} & {\color[HTML]{9A0000} \sig{.3395}} & -0.04 & {\color[HTML]{9A0000} \sig{-0.61}} & {\color[HTML]{9A0000} \sig{-1.37}} & {\color[HTML]{9A0000} \sig{-2.16}} & {\color[HTML]{9A0000} \sig{-4.05}} & {\color[HTML]{9A0000} \sig{.3581}} & +0.01 & -0.16 & {\color[HTML]{9A0000} \sig{-0.52}} & {\color[HTML]{9A0000} \sig{-0.89}} & {\color[HTML]{9A0000} \sig{-1.65}} \\
 & x10 & {\color[HTML]{9A0000} \sig{.5437}} & -0.84                              & {\color[HTML]{9A0000} \sig{-1.42}} & {\color[HTML]{9A0000} \sig{-1.63}} & {\color[HTML]{9A0000} \sig{-1.87}} & {\color[HTML]{9A0000} \sig{-2.31}} & {\color[HTML]{9A0000} \sig{.6734}} & {\color[HTML]{9A0000} \sig{-2.23}} & {\color[HTML]{9A0000} \sig{-2.4}}  & {\color[HTML]{9A0000} \sig{-2.23}} & {\color[HTML]{9A0000} \sig{-2.14}} & {\color[HTML]{9A0000} \sig{-2.16}} & {\color[HTML]{9A0000} \sig{.3368}} & 0     & {\color[HTML]{9A0000} \sig{-0.86}} & {\color[HTML]{9A0000} \sig{-2.03}} & {\color[HTML]{9A0000} \sig{-3.18}} & {\color[HTML]{9A0000} \sig{-5.71}} & {\color[HTML]{9A0000} \sig{.3577}} & +0.26 & -0.08 & {\color[HTML]{9A0000} \sig{-0.53}} & {\color[HTML]{9A0000} \sig{-1.14}} & {\color[HTML]{9A0000} \sig{-2.11}} \\
 & x20 & {\color[HTML]{9A0000} \sig{.541}}  & -1.19                              & {\color[HTML]{9A0000} \sig{-1.83}} & {\color[HTML]{9A0000} \sig{-2.12}} & {\color[HTML]{9A0000} \sig{-2.35}} & {\color[HTML]{9A0000} \sig{-2.96}} & {\color[HTML]{9A0000} \sig{.6666}} & {\color[HTML]{9A0000} \sig{-3.42}} & {\color[HTML]{9A0000} \sig{-3.54}} & {\color[HTML]{9A0000} \sig{-3.17}} & {\color[HTML]{9A0000} \sig{-2.95}} & {\color[HTML]{9A0000} \sig{-3.06}} & {\color[HTML]{9A0000} \sig{.3347}} & -0.05 & {\color[HTML]{9A0000} \sig{-1.15}} & {\color[HTML]{9A0000} \sig{-2.52}} & {\color[HTML]{9A0000} \sig{-4.01}} & {\color[HTML]{9A0000} \sig{-6.97}} & {\color[HTML]{9A0000} \sig{.3576}} & +0.24 & -0.11 & {\color[HTML]{9A0000} \sig{-0.57}} & {\color[HTML]{9A0000} \sig{-1.16}} & {\color[HTML]{9A0000} \sig{-2.14}} \\
 & x50 & {\color[HTML]{9A0000} \sig{.5376}} & {\color[HTML]{9A0000} \sig{-1.49}} & {\color[HTML]{9A0000} \sig{-2.35}} & {\color[HTML]{9A0000} \sig{-2.71}} & {\color[HTML]{9A0000} \sig{-3.02}} & {\color[HTML]{9A0000} \sig{-3.84}} & {\color[HTML]{9A0000} \sig{.6588}} & {\color[HTML]{9A0000} \sig{-4.71}} & {\color[HTML]{9A0000} \sig{-4.74}} & {\color[HTML]{9A0000} \sig{-4.31}} & {\color[HTML]{9A0000} \sig{-3.92}} & {\color[HTML]{9A0000} \sig{-4.2}}  & {\color[HTML]{9A0000} \sig{.3328}} & -0.17 & {\color[HTML]{9A0000} \sig{-1.42}} & {\color[HTML]{9A0000} \sig{-3.01}} & {\color[HTML]{9A0000} \sig{-4.67}} & {\color[HTML]{9A0000} \sig{-8}}    & {\color[HTML]{9A0000} \sig{.3576}} & +0.24 & -0.09 & {\color[HTML]{9A0000} \sig{-0.58}} & {\color[HTML]{9A0000} \sig{-1.18}} & {\color[HTML]{9A0000} \sig{-2.16}} \\
 & x80 & {\color[HTML]{9A0000} \sig{.5359}} & {\color[HTML]{9A0000} \sig{-1.76}} & {\color[HTML]{9A0000} \sig{-2.68}} & {\color[HTML]{9A0000} \sig{-2.95}} & {\color[HTML]{9A0000} \sig{-3.32}} & {\color[HTML]{9A0000} \sig{-4.19}} & {\color[HTML]{9A0000} \sig{.6548}} & {\color[HTML]{9A0000} \sig{-5.34}} & {\color[HTML]{9A0000} \sig{-5.29}} & {\color[HTML]{9A0000} \sig{-4.83}} & {\color[HTML]{9A0000} \sig{-4.46}} & {\color[HTML]{9A0000} \sig{-4.91}} & {\color[HTML]{9A0000} \sig{.3316}} & +0.03 & {\color[HTML]{9A0000} \sig{-1.52}} & {\color[HTML]{9A0000} \sig{-3.34}} & {\color[HTML]{9A0000} \sig{-5.18}} & {\color[HTML]{9A0000} \sig{-8.87}} & {\color[HTML]{9A0000} \sig{.3576}} & +0.23 & -0.1  & {\color[HTML]{9A0000} \sig{-0.59}} & {\color[HTML]{9A0000} \sig{-1.18}} & {\color[HTML]{9A0000} \sig{-2.17}} \\ \hline

\multirow{5}{*}{\rotatebox[origin=c]{90}{Util}} & x5  & {\color[HTML]{009901} \sig{.5567}} & {\color[HTML]{009901} \sig{+1.5}}  & +0.99                              & +0.53                              & +0.32                              & +0.26                              & {\color[HTML]{9A0000} \sig{.6846}} & -0.46                              & -0.5                               & -0.51                              & {\color[HTML]{9A0000} \sig{-0.66}} & {\color[HTML]{9A0000} \sig{-0.72}} & 0.3454                             & +0.12 & +0.11                              & -0.16                              & +0.07                              & -0.3                               & 0.3607                             & -0.01 & +0.06 & 0                                  & -0.04                              & +0.12                              \\
 & x10 & {\color[HTML]{009901} \sig{.5574}} & {\color[HTML]{009901} \sig{+2.05}} & {\color[HTML]{009901} \sig{+1.24}} & +0.67                              & +0.33                              & +0.2                               & {\color[HTML]{9A0000} \sig{.6807}} & -1.11                              & {\color[HTML]{9A0000} \sig{-1.16}} & {\color[HTML]{9A0000} \sig{-1.1}}  & {\color[HTML]{9A0000} \sig{-1.19}} & {\color[HTML]{9A0000} \sig{-1.22}} & 0.3453                             & +0.13 & +0.06                              & -0.11                              & +0                                 & -0.43                              & 0.3607                             & 0     & +0.12 & +0.01                              & -0.06                              & +0                                 \\
 & x20 & {\color[HTML]{009901} \sig{.5579}} & {\color[HTML]{009901} \sig{+2.52}} & {\color[HTML]{009901} \sig{+1.59}} & +0.71                              & +0.27                              & +0.08                              & {\color[HTML]{9A0000} \sig{.6762}} & {\color[HTML]{9A0000} \sig{-1.85}} & {\color[HTML]{9A0000} \sig{-2.03}} & {\color[HTML]{9A0000} \sig{-1.79}} & {\color[HTML]{9A0000} \sig{-1.73}} & {\color[HTML]{9A0000} \sig{-1.72}} & 0.3454                             & +0.11 & +0.06                              & -0.06                              & +0.15                              & -0.5                               & 0.3607                             & 0     & +0.14 & +0.05                              & -0.08                              & +0.01                              \\
 & x50 & {\color[HTML]{009901} \sig{.5579}} & {\color[HTML]{009901} \sig{+2.92}} & {\color[HTML]{009901} \sig{+1.8}}  & +0.64                              & +0.12                              & -0.16                              & {\color[HTML]{9A0000} \sig{.6722}} & {\color[HTML]{9A0000} \sig{-2.66}} & {\color[HTML]{9A0000} \sig{-2.85}} & {\color[HTML]{9A0000} \sig{-2.53}} & {\color[HTML]{9A0000} \sig{-2.13}} & {\color[HTML]{9A0000} \sig{-2.05}} & 0.3458                             & +0.19 & +0.07                              & +0.13                              & +0.22                              & -0.3                               & 0.3608                             & +0.16 & +0.23 & -0.03                              & -0.05                              & -0.04                              \\
 & x80 & {\color[HTML]{009901} \sig{.5577}} & {\color[HTML]{009901} \sig{+3.12}} & {\color[HTML]{009901} \sig{+1.86}} & +0.61                              & +0.01                              & -0.33                              & {\color[HTML]{9A0000} \sig{.6715}} & {\color[HTML]{9A0000} \sig{-2.92}} & {\color[HTML]{9A0000} \sig{-3}}    & {\color[HTML]{9A0000} \sig{-2.65}} & {\color[HTML]{9A0000} \sig{-2.15}} & {\color[HTML]{9A0000} \sig{-2.05}} & 0.346                              & +0.23 & +0.09                              & +0.15                              & +0.29                              & -0.18                              & 0.3608                             & +0.16 & +0.21 & +0.01                              & -0.02                              & -0.06                             
 \\ \hline
\end{tabular}
\end{table*}

\begin{figure}[t]
    \centering 
    \includegraphics[width=\columnwidth]{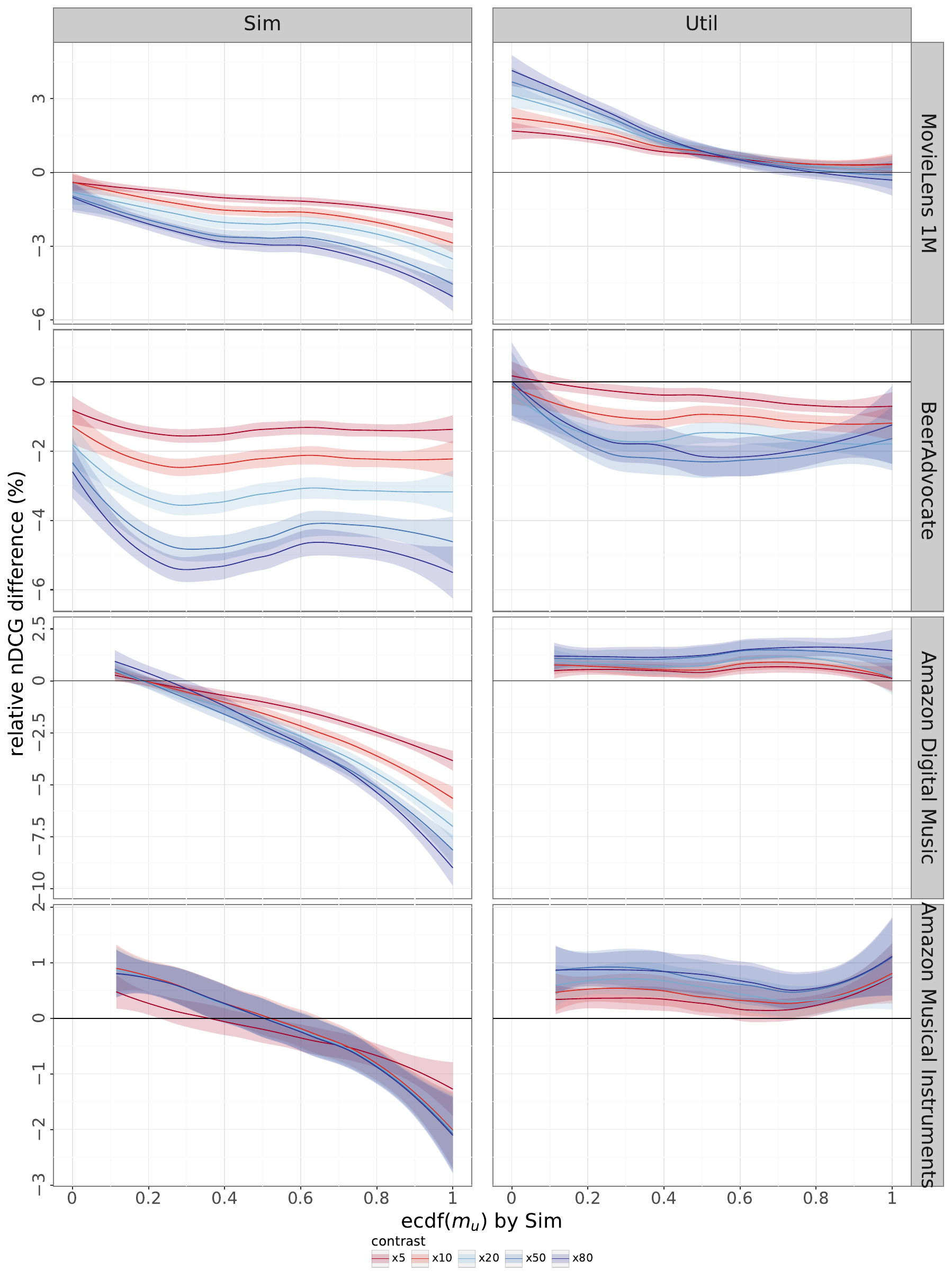}
    \caption{Same as Fig.~\ref{fig:res_csl_by_ndcg}, but plotted against ecdf($m_u$) by \Sim instead of ecdf(test nDCG) in the baseline model.}
    \label{fig:res_csl_by_ms}
\end{figure}

An effective mitigation of the mainstream bias would be reflected in increased performance for the lower groups (i.e. mainly `low' and `med-low'), ideally with no detriment to the higher groups and, especially, overall. In the previous section we separated users into groups by each of \Sim and \Util, but here we separate them directly by their test $nDCG$ with the baseline model FM, because this better illustrates how non-mainstream users suffer from the bias.

Table~\ref{tab:overall_performance} reports the relative percentage improvement in $nDCG$ scores per user group, as well as the overall mean score across all users in the dataset. We can clearly see that the use of \Sim benefits the non-mainstream users only in the two Amazon dataset; in MovieLens and BeerAdvocate they are even hurt further.
In contrast, \Util is always able to improve the utility of non-mainstream users across datasets, achieving relative $nDCG$ improvements of up to $5\%$ in the Amazon datasets. Improvements on the lower user groups are generally higher than losses on the higher groups, where users already receive (very) high recommendation utility anyway and such minor losses are probably unnoticed. This redistribution of model performance has a negligible effect on the global performance of the models, as evidenced by the overall $nDCG$ scores. This means that, with proper selection of the contrast in the cost function, \Util can minimize the mainstream bias at virtually no overall loss in utility. On the other hand, the use of \Sim for training leads to inferior overall performance on all four datasets.

Fig.~\ref{fig:res_csl_by_ndcg} presents a more fine-grained picture with one of the three random initializations in our experiments. Curve segments above 0 represent an improvement by the cost-sensitive models, while segments below 0 represent a loss. We can confirm that the cost-sensitive approach indeed makes the models focus on the non-mainstream users, as shown by the nicely smooth correlation between observed utility and relative improvement, moderated by the contrast in the cost function.
As expected though, this focus on the non-mainstream users comes at the cost of a utility loss for the mainstream users on the right-hand side of the plots. Nevertheless, when using \Util the relative loss for those users is generally much smaller than the gain for the very non-mainstream users, which are our main target.
The figure also shows that the actual relation between improvement and utility varies across datasets, as reflected by the different curve shapes. This is explained by the differences in the shape of their $nDCG$ distributions (see Fig.~\ref{fig:simvsutil}); recall that we use the $ecdf$ of the scores.
In a side-by-side comparison between \Sim and \Util, we see that \Util offers better performance nearly everywhere along the $x$-axis, but especially for the non-mainstream users. 

In summary, we see that our cost-sensitive approach brings better balance across users, thus helping in the mitigation of the mainstream bias. In addition, we confirm that an implicit quantification of mainstreamness like \Util works better than an explicit quantification like \Sim in steering the learning process towards better recommendations for the users that receive low utility from the baseline model. In addition, we note that the mitigation effect via \Util does not decay with increasing data sparsity (refer back to Table~\ref{tab:stats}).

One could be tempted to argue that \Util should obviously offer better results than \Sim when analyzing \emph{test} $nDCG$ because it is based on \emph{validation} $nDCG$ scores; test and validation scores should be highly correlated (we will come back to this in Section~\ref{sec:discussion}). After all, both Table~\ref{tab:overall_performance} and Fig.~\ref{fig:res_csl_by_ndcg} analyze results by test $nDCG$. The argument made above is that differences between mainstream and non-mainstream users can be immediately identified by test scores, but for the sake of clarity and to avoid potentially unfair assessment towards \Sim, Table~\ref{tab:overall_performance_by_ms} reports the same results but separating users by \Sim, while Fig.~\ref{fig:res_csl_by_ms} does so by plotting against \Sim.
While the results are less clear with this partition of users, the table confirms that models trained with \Sim are generally better at mitigating the bias than those trained with \Util. In particular, results for the BeerAdvocate dataset show that higher contrasts even lead to worse performance for the lower user groups, suggesting that \Sim is perhaps not properly identifying non-mainstream users.
The figure shows that \Util improves over the baseline across all levels of mainstreamness in the Amazon datasets, further suggesting that \Sim identifies as non-mainstream users that are probably not.
In summary, and even though this comparison could in turn be considered favorable to \Sim (note that previously we assessed against \emph{test} $nDCG$, not against the \emph{validation} $nDCG$ calculated by \Util), the results again support the use of \Util to quantify user mainstreamness and mitigate the bias. 

\section{Discussion} \label{sec:discussion}

A key assumption of our approach based on \Util is that we can reliably use utility, measured as the accuracy on a validation set, to determine the weight that each user should have in the training process. This implies that the accuracy on the validation set is a good estimate of the accuracy on the test set, which is where the effect will ultimately be assessed. If there was a low correlation between validation and test accuracy, the loss function would apply high weights for users that do not really need it, limiting or even altogether canceling the potential of our approach.

Intuitively, how well validation and test scores correlate is mainly determined by the amount of data. If only a few interactions are involved in the calculation of accuracy, the resulting scores will bear a high degree of noise or random error, thus lowering the correlation. In principle, we would therefore use as much data as possible in the validation and test sets. However, we would generally prefer to use all that data to actually train the model, but we note that the validation scores are somehow part of the training process itself, because they determine the weights.

A balance is therefore necessary, so we need to study the strength of the validation-test scores correlation as a function of the number of interactions in their data partitions. We did this by running the baseline FM model on different data partitions with varying minimum numbers of relevant items in the training set (3, 4, 5 and 10), and validation and test sets (1, 2, 3, 4 and 5 each). The actual split was conducted maintaining proportions (i.e. for the combination of 4/3/3 minimum items per set, a user has 40\% of their relevant items for training, 30\% for validation, and 30\% for testing).
We then measured the strength of the validation-test correlation via the RMSE of the scores and their Spearman $\rho$ correlation.

\begin{figure}
    \centering 
    \includegraphics[width=.45\textwidth]{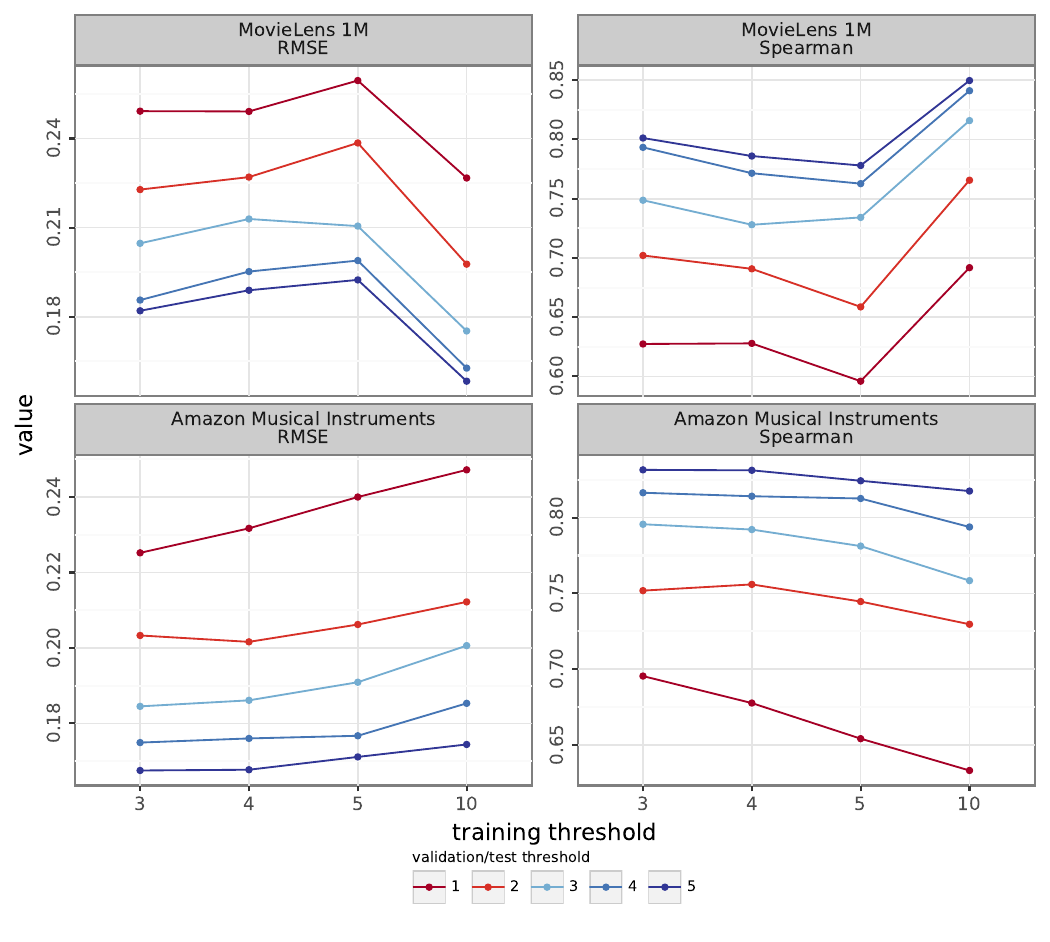}
    \caption{Correlation between validation and test scores as a function of the amount of data used for training, validation, and testing, for two sample datasets (most and least dense).}
    \label{fig:corr}
\end{figure}
Fig.~\ref{fig:corr} shows that, as expected, the correlation increases (low RMSE, high $\rho$) with the number of relevant interactions used in the validation and test sets. More interestingly, it shows that the amount of training data has a much smaller and varying effect, so despite it being a major factor to maximize model performance, it is not so to robustly assess that performance.
The plots indicate that requiring only one or two interactions in the validation set would lead to noisy scores; four interactions seem the bare minimum. As for the training set, the usual practice of having at least as much data as for validation and testing still applies in this context of non-time-aware recommendation.

All in all, our suggestion for this line of research on mainstream bias that works at the individual user level, is to have no less than four items per user in each of the three standard data partitions. Because the strength of the correlation is a key factor in our approach, we decided to require at least five to be on the safe side.
In fact, we also observed that the effect of cost-sensitive learning in the validation sets is similar to what is reported in Figs.~\ref{fig:res_csl_by_ndcg} and ~\ref{fig:res_csl_by_ms}.


\section{Conclusions and Future Work} \label{sec:con}

In this paper, we tackled the challenge of mainstream bias in CF-based recommendation. The main aspect we focused on is to steer the process of mitigating this bias directly by the utility resulting from the recommendation model and data at hand. For this purpose, we proposed an approach that assigns each user an importance weight during training, with these weights defined in a cost-sensitive manner.
By choosing to steer the model directly towards the users that receive low utility, and not towards those that \emph{appear} to be non-mainstream, we avoid the model to focus on users that already receive high utility even if they were not expected to. This way, the model does focus on the niche users that suffer from the bias.

Empirical results show that such models produce a more effective balance of the recommendation utility among the mainstream and non-mainstream users, in a way that is consistent across datasets with varying properties.
In addition, we provide suggestions regarding the minimum number of interactions to require when partitioning datasets. Without enough interactions, research on mainstream bias at the level of individual users might produce unreliable results.

For future work, we will first explore other ways to quantify mainstreamness. In the implicit measurement sense, an evident question is whether other metrics such as $AP$, or even the combination of multiple metrics, work better at identifying niche users. Additionally, we can think of ways to make the validation-test correlation robust to issues like sample selection bias, for example via inverse propensity scoring.
Another line is to explore more principled approaches for an explicit quantification through an extensive study of the factors that influence mainstreamness, such as the temporal dynamics.

Regarding our cost-sensitive learning approach, we will explore its generality, to see how it works for underlying models other than FM or other ranking frameworks such as pairwise and listwise.
We will also investigate the combination of cost-sensitive and adversarial learning strategies to mitigate mainstream bias: cost-sensitive to tell the model where to focus on, and adversary to tell how.

Finally, we note that our focus in this paper has been on the effect of mainstream
bias mitigation on the users, but one could wonder about what effect it has on the items. One hypothesis is that non-mainstream users are better served because the less popular items are now more likely to be recommended, so it would be interesting to study whether mitigating one bias amplifies or mitigates other biases, such as popularity or position.

\bibliographystyle{ACM-Reference-Format}
\balance
\bibliography{sample-base}


\begin{thebibliography}{24}


\ifx \showCODEN    \undefined \def \showCODEN     #1{\unskip}     \fi
\ifx \showDOI      \undefined \def \showDOI       #1{#1}\fi
\ifx \showISBNx    \undefined \def \showISBNx     #1{\unskip}     \fi
\ifx \showISBNxiii \undefined \def \showISBNxiii  #1{\unskip}     \fi
\ifx \showISSN     \undefined \def \showISSN      #1{\unskip}     \fi
\ifx \showLCCN     \undefined \def \showLCCN      #1{\unskip}     \fi
\ifx \shownote     \undefined \def \shownote      #1{#1}          \fi
\ifx \showarticletitle \undefined \def \showarticletitle #1{#1}   \fi
\ifx \showURL      \undefined \def \showURL       {\relax}        \fi
\providecommand\bibfield[2]{#2}
\providecommand\bibinfo[2]{#2}
\providecommand\natexlab[1]{#1}
\providecommand\showeprint[2][]{arXiv:#2}

\bibitem[Bauer and Schedl(2018)]%
        {DBLP:conf/ismir/0001S18}
\bibfield{author}{\bibinfo{person}{Christine Bauer} {and}
  \bibinfo{person}{Markus Schedl}.} \bibinfo{year}{2018}\natexlab{}.
\newblock \showarticletitle{Investigating Cross-Country Relationship between
  Users' Social Ties and Music Mainstreaminess}. In
  \bibinfo{booktitle}{\emph{{ISMIR}}}. \bibinfo{pages}{678--686}.
\newblock


\bibitem[Breunig et~al\mbox{.}(2000)]%
        {DBLP:conf/sigmod/BreunigKNS00}
\bibfield{author}{\bibinfo{person}{Markus~M. Breunig},
  \bibinfo{person}{Hans{-}Peter Kriegel}, \bibinfo{person}{Raymond~T. Ng},
  {and} \bibinfo{person}{J{\"{o}}rg Sander}.} \bibinfo{year}{2000}\natexlab{}.
\newblock \showarticletitle{{LOF:} Identifying Density-Based Local Outliers}.
  In \bibinfo{booktitle}{\emph{{SIGMOD} Conference}}.
  \bibinfo{publisher}{{ACM}}, \bibinfo{pages}{93--104}.
\newblock


\bibitem[Ekstrand(2020)]%
        {DBLP:conf/cikm/Ekstrand20}
\bibfield{author}{\bibinfo{person}{Michael~D. Ekstrand}.}
  \bibinfo{year}{2020}\natexlab{}.
\newblock \showarticletitle{LensKit for Python: Next-Generation Software for
  Recommender Systems Experiments}. In \bibinfo{booktitle}{\emph{{CIKM}}}.
  \bibinfo{publisher}{{ACM}}, \bibinfo{pages}{2999--3006}.
\newblock


\bibitem[Ekstrand et~al\mbox{.}(2018)]%
        {DBLP:conf/fat/EkstrandTAEAMP18}
\bibfield{author}{\bibinfo{person}{Michael~D. Ekstrand}, \bibinfo{person}{Mucun
  Tian}, \bibinfo{person}{Ion~Madrazo Azpiazu}, \bibinfo{person}{Jennifer~D.
  Ekstrand}, \bibinfo{person}{Oghenemaro Anuyah}, \bibinfo{person}{David
  McNeill}, {and} \bibinfo{person}{Maria~Soledad Pera}.}
  \bibinfo{year}{2018}\natexlab{}.
\newblock \showarticletitle{All The Cool Kids, How Do They Fit In?: Popularity
  and Demographic Biases in Recommender Evaluation and Effectiveness}. In
  \bibinfo{booktitle}{\emph{{FAT}}} \emph{(\bibinfo{series}{Proceedings of
  Machine Learning Research}, Vol.~\bibinfo{volume}{81})}.
  \bibinfo{publisher}{{PMLR}}, \bibinfo{pages}{172--186}.
\newblock


\bibitem[Goldberg et~al\mbox{.}(1992)]%
        {DBLP:journals/cacm/GoldbergNOT92}
\bibfield{author}{\bibinfo{person}{David Goldberg}, \bibinfo{person}{David~A.
  Nichols}, \bibinfo{person}{Brian~M. Oki}, {and} \bibinfo{person}{Douglas~B.
  Terry}.} \bibinfo{year}{1992}\natexlab{}.
\newblock \showarticletitle{Using Collaborative Filtering to Weave an
  Information Tapestry}.
\newblock \bibinfo{journal}{\emph{Commun. {ACM}}} \bibinfo{volume}{35},
  \bibinfo{number}{12} (\bibinfo{year}{1992}), \bibinfo{pages}{61--70}.
\newblock


\bibitem[Harper and Konstan(2016)]%
        {DBLP:journals/tiis/HarperK16}
\bibfield{author}{\bibinfo{person}{F.~Maxwell Harper} {and}
  \bibinfo{person}{Joseph~A. Konstan}.} \bibinfo{year}{2016}\natexlab{}.
\newblock \showarticletitle{The MovieLens Datasets: History and Context}.
\newblock \bibinfo{journal}{\emph{{ACM} Trans. Interact. Intell. Syst.}}
  \bibinfo{volume}{5}, \bibinfo{number}{4} (\bibinfo{year}{2016}),
  \bibinfo{pages}{19:1--19:19}.
\newblock


\bibitem[He et~al\mbox{.}(2017)]%
        {DBLP:conf/www/HeLZNHC17}
\bibfield{author}{\bibinfo{person}{Xiangnan He}, \bibinfo{person}{Lizi Liao},
  \bibinfo{person}{Hanwang Zhang}, \bibinfo{person}{Liqiang Nie},
  \bibinfo{person}{Xia Hu}, {and} \bibinfo{person}{Tat{-}Seng Chua}.}
  \bibinfo{year}{2017}\natexlab{}.
\newblock \showarticletitle{Neural Collaborative Filtering}. In
  \bibinfo{booktitle}{\emph{{WWW}}}. \bibinfo{publisher}{{ACM}},
  \bibinfo{pages}{173--182}.
\newblock


\bibitem[Kingma and Ba(2015)]%
        {DBLP:journals/corr/KingmaB14}
\bibfield{author}{\bibinfo{person}{Diederik~P. Kingma} {and}
  \bibinfo{person}{Jimmy Ba}.} \bibinfo{year}{2015}\natexlab{}.
\newblock \showarticletitle{Adam: {A} Method for Stochastic Optimization}. In
  \bibinfo{booktitle}{\emph{{ICLR} (Poster)}}.
\newblock


\bibitem[Leonhardt et~al\mbox{.}(2018)]%
        {DBLP:conf/www/LeonhardtAK18}
\bibfield{author}{\bibinfo{person}{Jurek Leonhardt}, \bibinfo{person}{Avishek
  Anand}, {and} \bibinfo{person}{Megha Khosla}.}
  \bibinfo{year}{2018}\natexlab{}.
\newblock \showarticletitle{User Fairness in Recommender Systems}. In
  \bibinfo{booktitle}{\emph{{WWW} (Companion Volume)}}.
  \bibinfo{publisher}{{ACM}}, \bibinfo{pages}{101--102}.
\newblock


\bibitem[Li et~al\mbox{.}(2021b)]%
        {DBLP:conf/wsdm/LiUH21}
\bibfield{author}{\bibinfo{person}{Roger~Zhe Li}, \bibinfo{person}{Juli{\'{a}}n
  Urbano}, {and} \bibinfo{person}{Alan Hanjalic}.}
  \bibinfo{year}{2021}\natexlab{b}.
\newblock \showarticletitle{Leave No User Behind: Towards Improving the Utility
  of Recommender Systems for Non-mainstream Users}. In
  \bibinfo{booktitle}{\emph{{WSDM}}}. \bibinfo{publisher}{{ACM}},
  \bibinfo{pages}{103--111}.
\newblock


\bibitem[Li et~al\mbox{.}(2021a)]%
        {DBLP:conf/www/LiCFGZ21}
\bibfield{author}{\bibinfo{person}{Yunqi Li}, \bibinfo{person}{Hanxiong Chen},
  \bibinfo{person}{Zuohui Fu}, \bibinfo{person}{Yingqiang Ge}, {and}
  \bibinfo{person}{Yongfeng Zhang}.} \bibinfo{year}{2021}\natexlab{a}.
\newblock \showarticletitle{User-oriented Fairness in Recommendation}. In
  \bibinfo{booktitle}{\emph{{WWW}}}. \bibinfo{publisher}{{ACM} / {IW3C2}},
  \bibinfo{pages}{624--632}.
\newblock


\bibitem[Liu et~al\mbox{.}(2019)]%
        {DBLP:conf/kdd/LiuGZL19}
\bibfield{author}{\bibinfo{person}{Yudan Liu}, \bibinfo{person}{Kaikai Ge},
  \bibinfo{person}{Xu Zhang}, {and} \bibinfo{person}{Leyu Lin}.}
  \bibinfo{year}{2019}\natexlab{}.
\newblock \showarticletitle{Real-time Attention Based Look-alike Model for
  Recommender System}. In \bibinfo{booktitle}{\emph{{KDD}}}.
  \bibinfo{publisher}{{ACM}}, \bibinfo{pages}{2765--2773}.
\newblock


\bibitem[McAuley et~al\mbox{.}(2012)]%
        {DBLP:conf/icdm/McAuleyLJ12}
\bibfield{author}{\bibinfo{person}{Julian~J. McAuley}, \bibinfo{person}{Jure
  Leskovec}, {and} \bibinfo{person}{Dan Jurafsky}.}
  \bibinfo{year}{2012}\natexlab{}.
\newblock \showarticletitle{Learning Attitudes and Attributes from Multi-aspect
  Reviews}. In \bibinfo{booktitle}{\emph{{ICDM}}}. \bibinfo{publisher}{{IEEE}
  Computer Society}, \bibinfo{pages}{1020--1025}.
\newblock


\bibitem[Melchiorre et~al\mbox{.}(2021)]%
        {DBLP:journals/ipm/MelchiorreRPBLS21}
\bibfield{author}{\bibinfo{person}{Alessandro~B. Melchiorre},
  \bibinfo{person}{Navid Rekabsaz}, \bibinfo{person}{Emilia
  Parada{-}Cabaleiro}, \bibinfo{person}{Stefan Brandl}, \bibinfo{person}{Oleg
  Lesota}, {and} \bibinfo{person}{Markus Schedl}.}
  \bibinfo{year}{2021}\natexlab{}.
\newblock \showarticletitle{Investigating gender fairness of recommendation
  algorithms in the music domain}.
\newblock \bibinfo{journal}{\emph{Inf. Process. Manag.}} \bibinfo{volume}{58},
  \bibinfo{number}{5} (\bibinfo{year}{2021}), \bibinfo{pages}{102666}.
\newblock


\bibitem[Naghiaei et~al\mbox{.}(2022)]%
        {DBLP:conf/sigir/NaghiaeiRD22}
\bibfield{author}{\bibinfo{person}{Mohammadmehdi Naghiaei},
  \bibinfo{person}{Hossein~A. Rahmani}, {and} \bibinfo{person}{Yashar
  Deldjoo}.} \bibinfo{year}{2022}\natexlab{}.
\newblock \showarticletitle{CPFair: Personalized Consumer and Producer Fairness
  Re-ranking for Recommender Systems}. In \bibinfo{booktitle}{\emph{{SIGIR}}}.
  \bibinfo{publisher}{{ACM}}, \bibinfo{pages}{770--779}.
\newblock


\bibitem[Ni et~al\mbox{.}(2019)]%
        {DBLP:conf/emnlp/NiLM19}
\bibfield{author}{\bibinfo{person}{Jianmo Ni}, \bibinfo{person}{Jiacheng Li},
  {and} \bibinfo{person}{Julian~J. McAuley}.} \bibinfo{year}{2019}\natexlab{}.
\newblock \showarticletitle{Justifying Recommendations using Distantly-Labeled
  Reviews and Fine-Grained Aspects}. In
  \bibinfo{booktitle}{\emph{{EMNLP/IJCNLP} {(1)}}}.
  \bibinfo{publisher}{Association for Computational Linguistics},
  \bibinfo{pages}{188--197}.
\newblock


\bibitem[Paszke et~al\mbox{.}(2019)]%
        {DBLP:conf/nips/PaszkeGMLBCKLGA19}
\bibfield{author}{\bibinfo{person}{Adam Paszke}, \bibinfo{person}{Sam Gross},
  \bibinfo{person}{Francisco Massa}, \bibinfo{person}{Adam Lerer},
  \bibinfo{person}{James Bradbury}, \bibinfo{person}{Gregory Chanan},
  \bibinfo{person}{Trevor Killeen}, \bibinfo{person}{Zeming Lin},
  \bibinfo{person}{Natalia Gimelshein}, \bibinfo{person}{Luca Antiga},
  \bibinfo{person}{Alban Desmaison}, \bibinfo{person}{Andreas K{\"{o}}pf},
  \bibinfo{person}{Edward~Z. Yang}, \bibinfo{person}{Zachary DeVito},
  \bibinfo{person}{Martin Raison}, \bibinfo{person}{Alykhan Tejani},
  \bibinfo{person}{Sasank Chilamkurthy}, \bibinfo{person}{Benoit Steiner},
  \bibinfo{person}{Lu Fang}, \bibinfo{person}{Junjie Bai}, {and}
  \bibinfo{person}{Soumith Chintala}.} \bibinfo{year}{2019}\natexlab{}.
\newblock \showarticletitle{PyTorch: An Imperative Style, High-Performance Deep
  Learning Library}. In \bibinfo{booktitle}{\emph{NeurIPS}}.
  \bibinfo{pages}{8024--8035}.
\newblock


\bibitem[Rendle(2010)]%
        {DBLP:conf/icdm/Rendle10}
\bibfield{author}{\bibinfo{person}{Steffen Rendle}.}
  \bibinfo{year}{2010}\natexlab{}.
\newblock \showarticletitle{Factorization Machines}. In
  \bibinfo{booktitle}{\emph{{ICDM}}}. \bibinfo{publisher}{{IEEE} Computer
  Society}, \bibinfo{pages}{995--1000}.
\newblock


\bibitem[Ruff et~al\mbox{.}(2018)]%
        {DBLP:conf/icml/RuffGDSVBMK18}
\bibfield{author}{\bibinfo{person}{Lukas Ruff}, \bibinfo{person}{Nico
  G{\"{o}}rnitz}, \bibinfo{person}{Lucas Deecke}, \bibinfo{person}{Shoaib~Ahmed
  Siddiqui}, \bibinfo{person}{Robert~A. Vandermeulen},
  \bibinfo{person}{Alexander Binder}, \bibinfo{person}{Emmanuel M{\"{u}}ller},
  {and} \bibinfo{person}{Marius Kloft}.} \bibinfo{year}{2018}\natexlab{}.
\newblock \showarticletitle{Deep One-Class Classification}. In
  \bibinfo{booktitle}{\emph{{ICML}}} \emph{(\bibinfo{series}{Proceedings of
  Machine Learning Research}, Vol.~\bibinfo{volume}{80})}.
  \bibinfo{publisher}{{PMLR}}, \bibinfo{pages}{4390--4399}.
\newblock


\bibitem[Rumelhart et~al\mbox{.}(1985)]%
        {rumelhart1985learning}
\bibfield{author}{\bibinfo{person}{David~E Rumelhart},
  \bibinfo{person}{Geoffrey~E Hinton}, {and} \bibinfo{person}{Ronald~J
  Williams}.} \bibinfo{year}{1985}\natexlab{}.
\newblock \bibinfo{booktitle}{\emph{Learning internal representations by error
  propagation}}.
\newblock \bibinfo{type}{{T}echnical {R}eport}.
  \bibinfo{institution}{California Univ San Diego La Jolla Inst for Cognitive
  Science}.
\newblock


\bibitem[Sun et~al\mbox{.}(2020)]%
        {DBLP:conf/recsys/SunY00Q0G20}
\bibfield{author}{\bibinfo{person}{Zhu Sun}, \bibinfo{person}{Di Yu},
  \bibinfo{person}{Hui Fang}, \bibinfo{person}{Jie Yang},
  \bibinfo{person}{Xinghua Qu}, \bibinfo{person}{Jie Zhang}, {and}
  \bibinfo{person}{Cong Geng}.} \bibinfo{year}{2020}\natexlab{}.
\newblock \showarticletitle{Are We Evaluating Rigorously? Benchmarking
  Recommendation for Reproducible Evaluation and Fair Comparison}. In
  \bibinfo{booktitle}{\emph{RecSys}}. \bibinfo{publisher}{{ACM}},
  \bibinfo{pages}{23--32}.
\newblock


\bibitem[Thai{-}Nghe et~al\mbox{.}(2010)]%
        {DBLP:conf/ijcnn/Thai-NgheGS10}
\bibfield{author}{\bibinfo{person}{Nguyen Thai{-}Nghe}, \bibinfo{person}{Zeno
  Gantner}, {and} \bibinfo{person}{Lars Schmidt{-}Thieme}.}
  \bibinfo{year}{2010}\natexlab{}.
\newblock \showarticletitle{Cost-sensitive learning methods for imbalanced
  data}. In \bibinfo{booktitle}{\emph{{IJCNN}}}. \bibinfo{publisher}{{IEEE}},
  \bibinfo{pages}{1--8}.
\newblock


\bibitem[Wu et~al\mbox{.}(2022)]%
        {DBLP:conf/ijcai/WuWH22}
\bibfield{author}{\bibinfo{person}{Chuhan Wu}, \bibinfo{person}{Fangzhao Wu},
  {and} \bibinfo{person}{Yongfeng Huang}.} \bibinfo{year}{2022}\natexlab{}.
\newblock \showarticletitle{Rethinking InfoNCE: How Many Negative Samples Do
  You Need?}. In \bibinfo{booktitle}{\emph{{IJCAI}}}.
  \bibinfo{publisher}{ijcai.org}, \bibinfo{pages}{2509--2515}.
\newblock


\bibitem[Zhu and Caverlee(2022)]%
        {DBLP:conf/wsdm/ZhuC22}
\bibfield{author}{\bibinfo{person}{Ziwei Zhu} {and} \bibinfo{person}{James
  Caverlee}.} \bibinfo{year}{2022}\natexlab{}.
\newblock \showarticletitle{Fighting Mainstream Bias in Recommender Systems via
  Local Fine Tuning}. In \bibinfo{booktitle}{\emph{{WSDM}}}.
  \bibinfo{publisher}{{ACM}}, \bibinfo{pages}{1497--1506}.
\newblock


\end{thebibliography}

\end{document}